\documentclass{aa}

\usepackage{kotex}
\usepackage{graphicx}
\usepackage{txfonts}
\begin{document}

   \title{Long-period radial velocity variations of nine M red giants}

   \subtitle{The detection of sub-stellar companions around HD~6860 and HD~112300}

   \author{
          Byeong-Cheol Lee (이병철)\inst{1,2},
          Hee-Jin Do (도희진)\inst{3},
          Myeong-Gu Park (박명구)\inst{3},
          Beomdu Lim (임범두)\inst{4},\\
          Yeon-Ho Choi (최연호)\inst{1,2},
          Jae-Rim Koo (구재림)\inst{4},
          Tae-Yang Bang (방태양)\inst{3},
          Hyeong-Ill Oh (오형일)\inst{3},\\
          Inwoo Han (한인우)\inst{1},       
                    and
          Heon-Young Chang (장헌영)\inst{3}
          }

   \institute{Korea Astronomy and Space Science Institute, 776,
            Daedeokdae-Ro, Youseong-Gu, Daejeon 34055, Korea\\
            \email{bclee@kasi.re.kr}
            \and
            Korea University of Science and Technology, Gajeong-ro Yuseong-gu, Daejeon 34113, Korea
            \and
            Department of Astronomy and Atmospheric Sciences, Kyungpook National University, Daegu 41566, Korea\\
             \email{taekwon@nate.com}
            \and
            Kongju National University, Gongjudaehak-ro 56, Gongju-si, Chungcheongnam-do 32588, Korea
             }
   \date{Received April 6, 2022; accepted xxxx xx, 2023}


  \abstract
   {Certain periodic variations of radial velocities (RV) of wobbling giants originate from exoplanets. Indeed, a number of exoplanets have been discovered around giant stars.}
   {The purpose of our study is to find low-amplitude and long-period RV variations around bright M (super) giants in the RGB (or AGB) stage, which are long-period variables (LPVs) or high-proper-motion (HPM) stars.
   }
   {High-resolution, fiber-fed Bohyunsan Observatory Echelle Spectrograph (BOES) at the Bohyunsan Optical Astronomy Observatory (BOAO) was used to record numerous spectra of nine giants. The observation period for the targets spans 16 years, from 2005 to 2022.
    } 
   {We found from the precise RV observations of nine M giants two sub-stellar companions, one with a 28.26$^{+2.05}_{-2.17}$ $M_{J}$ orbiting  period of 663.87$^{+4.61}_{-4.31}$ days at a distance of 2.03$^{+0.01}_{-0.01}$ AU (HD~6860) and the other, with a 15.83$^{+2.33}_{-2.74}$ $M_{J}$ orbiting  period of 466.63 $^{+1.47}_{-1.28}$ days at a distance of 1.33 $^{+0.08}_{-0.11}$ AU (HD~112300). Our estimate of the stellar parameters for HD~6860 makes it currently the largest star with a sub-stellar companion. We also found RV variations mimicking a planetary companion in HD~18884 and confirmed LPVs in two stars, HD~39801 and HD~42995.
   The RV variations of some stars seem to be associated with stellar activities rather than reflex orbital motion  due to their companions. Such variations are also detected even for HD~6860 and HD~112300, hosting sub-stellar companions.
   }  
   {}

   \authorrunning{Do et al.}
   \titlerunning{Long-period radial velocity variations of nine M red giants}
   \maketitle
%

\section{Introduction}

As its core hydrogen supply runs out, a star begins hydrogen thermonuclear fusion in the shell surrounding the central nucleus. During this process, stars are referred to as red giants, with radii ranging from tens to hundreds of times that of our Sun. Red giants emit relatively more light than the Sun despite their lower surface temperature. The brightness of a red giant branch (RGB) star is about 3,000 times that of the sun. The surface temperature of an M-type RGB star is 3,000 to 4,000 K and the radius about 200 times that of the sun. Asymptotic-giant-branch (AGB) stars have luminosities similar to those of the brighter RGB stars and are up to several times more luminous at the end of the thermal pulsing phase \citep{1993ApJ...418..457S}.

Another interesting aspect of these stars is the existence of variations in the brightness over long-period referring to various groups of cool luminous pulsating variable stars. These long-period variables (LPVs) include the Mira and semiregular types, slow irregular variables, and OGLE small-amplitude red giants (OSARGs). LPVs are seen in both giant and supergiant stars \citep{2009AcA....59..239S} with periods from around one hundred days to more than one thousand days. In some cases, the variations are too poorly defined to identify the period, although it is an open question as to whether they are truly non-periodic.
Most LPVs are thermally pulsing AGB stars with luminosities several thousand times that of the Sun. Some semiregular and irregular variables are less luminous giant stars, while others are more luminous supergiants, including some of the largest known stars, such as VY CMa.

Low-amplitude and long-period radial velocity (RV) variations are common among  evolved K, M giant stars \citep{1993ApJ...413..339H,2003A&A...403.1077K,2013A&A...549A...2L,2015A&A...580A..31H}.
Such variations appear on both long (hundreds of days) and short (hours to days) time scales. Short-term variations are most likely caused by Sun-like oscillations  excited by turbulent convection and by convective motions in the outer convective zone. Long-term variations can arise from either sub-stellar companions, surface inhomogeneities, pulsations, or other intrinsic stellar mechanisms.
For giant stars, the surface granulation size is linearly dependent on the stellar radius and mass \citep{2002AN....323..213F}. Thus, the atmosphere of red giants should exhibit a small number of large cells that may cause stochastic low-amplitude RV or light variations. These different time-scale variations are, in general, detected in K and M giant stars and may be associated with complex effects \citep{1999ASPC..185..193L}.

In 2003, we initiated a precise RV survey of nine M giants as part of an ongoing K giant exoplanet survey using the 1.8 m telescope at the Bohyunsan Optical Astronomy Observatory (BOAO). The goal of this paper is to interpret measured low-amplitude and long-period RVs for M giants. In Section 2, we describe our observations and data analysis details. In Section 3, the stellar characteristics of the host stars are presented. Orbital solutions and  possible origins of RV variations are presented in Sect. 4. The results of the RV variation measurements taken here are described  in Sect. 5.   Finally, in Section 6, we discuss and summarize our findings.


\begin{sidewaystable*}
\caption{Stellar parameters}\label{tab1}
\centering
\begin{tabular}{lccccccccccc}
\hline
Parameter & HD 6860 & HD 18884 & HD 39801 & HD 42995 & HD 44478 & HD 112300 & HD 146051 & HD 156014 & HD 183030 &  Ref.\\
    &($\beta$ And) &($\alpha$ Cet) &($\alpha$ Ori) &($\eta$ Gem) &($\mu$ Gem) &($\delta$ Vir) & ($\delta$ Oph) & ($\alpha$ Her) & ($\lambda$ UMi) & \\
\hline
\hline
Spectral type					& M0 III	& M1.5 IIa & M1 & M3 III & M3 III & M1 III & M0.5 III & M5 Ib-II &M1 III& (1) \\
                                &HPM \tablefootmark{a} &LPV\tablefootmark{b}&SRC\tablefootmark{c}&SRC &LPV &LPV &HPM &LPV &LPV\\
$\textit{$m_{v}$}$ (mag)		& 2.06 & 2.56 & 0.58 & 3.32 & 2.9 & 3.38  & 2.74 & 3.48 & 6.38 & (1)  \\
$\textit{B -- V}$ (mag)			& 1.58 & 1.66 & 1.77 & 1.59 & 1.63 & 1.58 &  1.59 & 1.44 & 2.35 & (1)  \\
distance (pc)                   &60.45 &76.29 &149.79 &115.02 &70.81 &60.81 &52.46 &107.52 &269.86 & (2) \\
${HIP_{scat}}$ (mag)			& 0.012	    & 0.012 & 0.091 & 0.059 & 0.037 & 0.022 & 0.090 & 0.122 & 0.030 & (1) \\
$\pi$ (mas)				& 23.3 $\pm$ 0.03 & -- & -- & -- & -- & 14.0 $\pm$ 0.6   & 0.5 $\pm$ 20.4 & 0.5 $\pm$ 9.9& 3.7 $\pm$ 0.1 &  (3) \\
                        & -- & 13.1 $\pm$ 0.4 & 6.6 $\pm$ 0.8 & 8.5 $\pm$ 1.2 & -- & -- & --  & -- & -- &  (4) \\
$T_{\rm{eff}}$ (K)				& 3802& 3890 & 3490 & 3571 & 3615 &3657 &3811 & 3269 & 3583 &  (5) \\
$\rm{[Fe/H]}$				& $-$0.03 & $-$0.03 & 0.09 & 0 &$-$0.03 & $-$0.06 & $-$0.04 & -- & --  & (2) \\
log $\it g$ (cgs)				& 0.541 &0.564 &$-$0.318 &0.147 &0.398 &0.8 &0.936 &$-$0.21 &0.625 & (5) \\
$\textit{$R_{\star}$}$ ($R_{\odot}$)	& 86.4  &87 &713.6 &157.2 &105 &65.8 &54.2 &377.3 &78.9 & (5)  \\
$\textit{$M_{\star}$}$ ($M_{\odot}$)  & --  &--&--&--&2.1&1.4$\pm$0.3&1.5&--&--& (6) \\
                                      & --&--&--&--&--&--&--&2.5&--& (7) \\
                                      & --&--&16.5--19&--&--&--&--&--&--& (8) \\
                                      & --&--&--&2.5&--&--&--&--&--& (9) \\
                                      &2.49&--&--&--&--&--&--&--&--& (10) \\
                                      &--&2.3$\pm$ 0.2&--&--&--&--&--&--&--& (11) \\
$\textit{$L_{\star}$}$ ($L_{\odot}$)  & 1674.88& -- & 36554.42 & -- & -- & 502.59 &  715 & 1284.06 &672.19 &  (2)   \\
                                      & 1403 &1588  & 67879  &3613 &1692 &697 &558 &14613 &921 &(5)   \\
$v_{\rm{rot}}$ sin $i$ (km s$^{-1}$)  & 6.0	 &7.1 & -- & -- & -- &8.1 &5.9 &-- &-- & (13) \\
$P_{\rm{rot}}$ / sin $i$ (days)		  & 7900 & 6250 &--&--& -- & 4200 &460&--&--&  (13)\\
\hline
\hline
\end{tabular}
\tablebib{(1)~\citet{1997yCat.1239....0E};(2)~\citet{2012AstL...38..331A};(3)~\citet{2018A&A...616A...1G};(4)~\citet{2007A&A...474..653V};
(5)~\citet{2017MNRAS.471..770M};(6)~\citet{2007IAUS..239..307T};(7)~\citet{2013AJ....146..148M};(8)~\citet{2020ApJ...902...63J};
(9)~\citet{1998A&A...330..225H};(10)~\citet{2011A&A...533A.107D};(11)~\citet{2006A&A...460..855W};(12)~\citet{2008AJ....135..209M};
(13)This work\\
(a) High-proper motion star (b) Long-period variables (c) Semi-regular variable in late-type giants
}
\end{sidewaystable*}

\section{Observations and data reduction\label{sec:star}}
\begin{figure}[t]
\centering
\includegraphics[width=10cm]{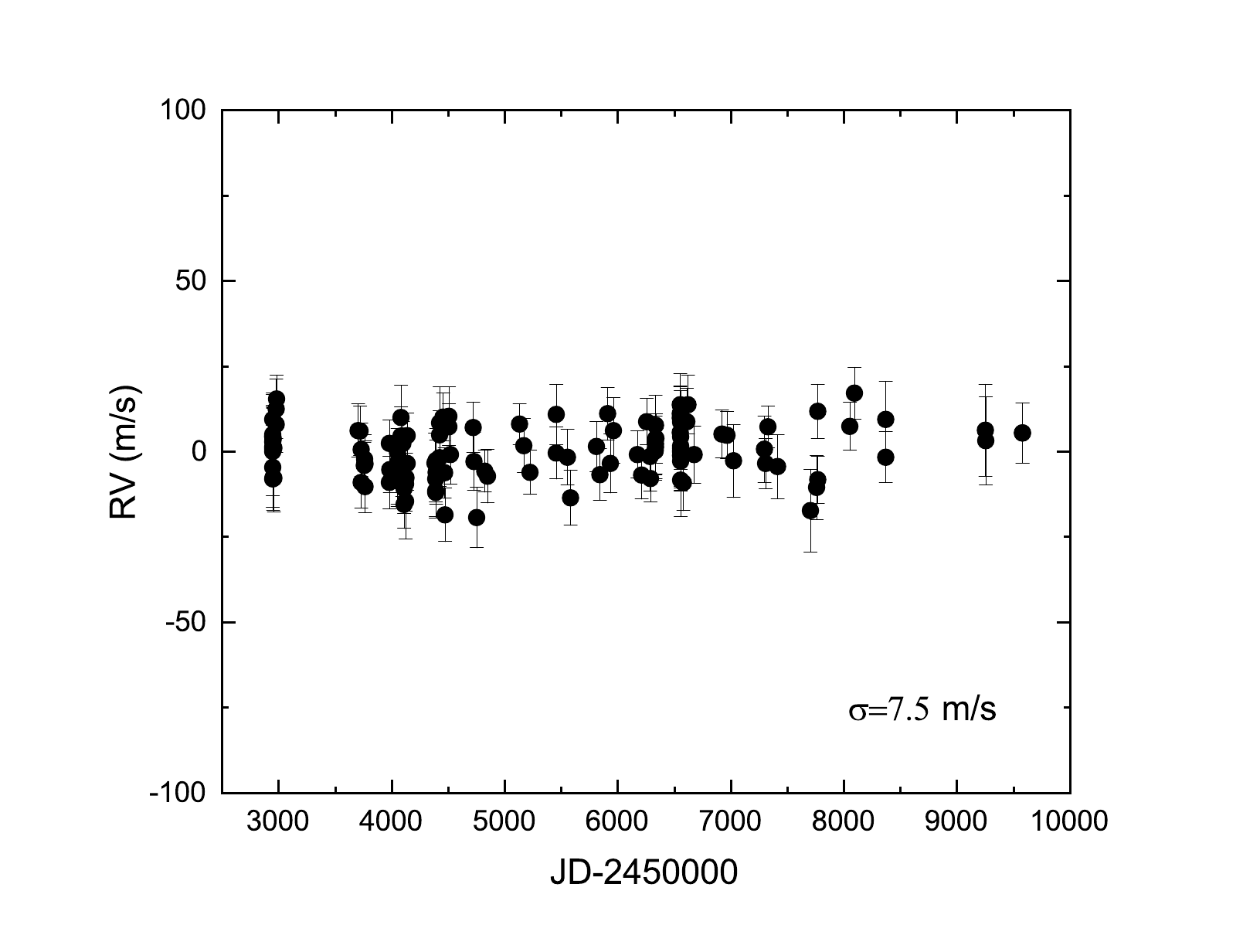}
\caption{Long-period RV variations for standard star $\tau$ Ceti from 2003 to 2021. \label{fig:std}}
\end{figure}

Observations were carried out using the fiber-fed high-resolution Bohyunsan Observatory Echelle Spectrograph (BOES) attached to the 1.8 m telescope at BOAO in Korea \citep{2007PASP..119.1052K}. One exposure with the BOES has a wavelength coverage range of 3500 ${\AA}$ to 10 500 ${\AA}$ distributed over approximately 75 spectral orders. The BOES is equipped with an iodine absorption (I$_{2}$) cell of the type needed for more precise RV measurements. Before starlight enters the fiber, it passes through the I$_{2}$ absorption cell regulated at 67\,$^{\circ}$C, which superimposes thousands of molecular absorption lines over the object spectra in the spectral region between 4900 and 6100 ${\AA}$. Using these lines as a wavelength standard, we simultaneously model the time-variant instrumental profile and Doppler shift relative to an I$_{2}$ free template spectrum.
In order to provide precise RV measurements, we used fiber with a diameter of 80 $\mu$m, which yields a  resolving power $\emph{R}$ = 90 000.
Most observations were conducted from 2005 to 2022.
 The estimated signal-to-noise (S/N) ratio in the I$_{2}$ region was approximately 200 with a typical exposure time ranging from 120 to 600 s.
The RV standard star $\tau$ Ceti, monitored since 2003, was shown to maintain a stable RV value of  7.5 m s$^{-1}$ for 17 years, as presented in Fig.~\ref{fig:std}

\begin{figure}[t]
\centering
\includegraphics[width=10cm]{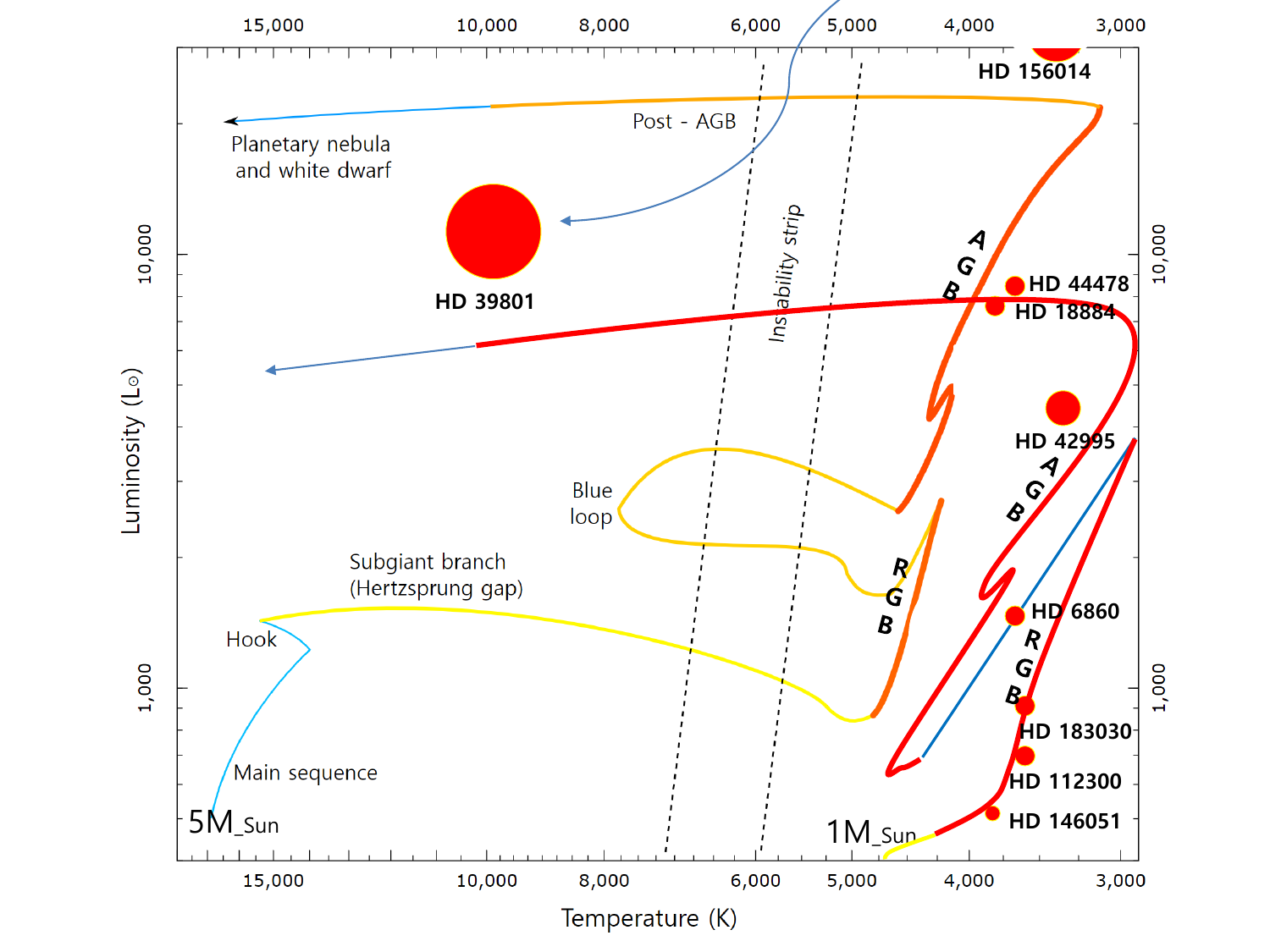}
\caption{Distribution in the H-R diagram of the data used for observations. \label{f2_HR}}
\end{figure}

Our stellar samples are the result of the intersection of several datasets. In this case, 1) all stars of spectral type M appearing in the \textit{HIPPARCOS} Catalogue \citep{1997yCat.1239....0E}, 2) stars from the northern hemisphere (δ > 30 degree) observed with the BOES at the BOAO, 3) all stars brighter than 6.5 (V magnitude) to attain a sufficient S/N, which are barely observable using  2-m class telescopes with a Doppler precision level of ∼10 m s$^{-1}$, 4) M giants showing high-proper motion, long-period variation or semi-normal variables.

The standard reduction procedures of flat-fielding, scattered light subtraction, and order extraction from raw CCD images were carried out using the IRAF software package. Precise RV measurements using the I$_{2}$ method were undertaken using RVI2CELL \citep{2007PKAS...22...75H}, which is based on a method used in \citet{1995PASP..107..966V, 1996PASP..108..500B}. However, for the modeling of the instrument profile, we used the matrix formula described by Endl et al. (2000).
We solved the matrix equation using singular value decomposition instead of using the maximum entropy method
\citep{2000A&A...362..585E}.

\section{Stellar Characteristics\label{sec:star}}

%
   \begin{figure*}
   \centering
   \includegraphics[width=21cm]{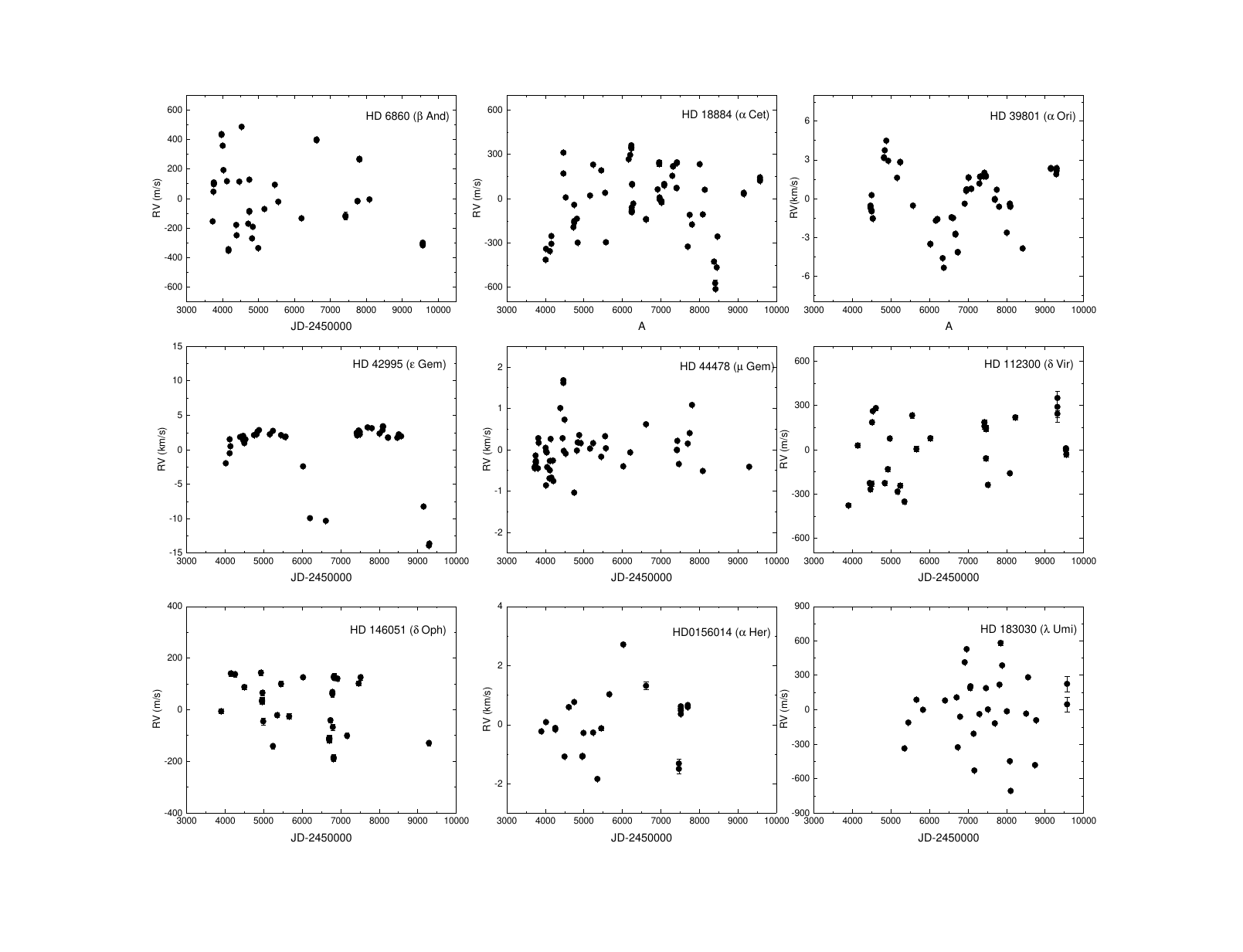}
      \caption{RV variations for nine M giants. The names of individual targets are labeled at the upper right corners of each panel.
              }
         \label{f3_all_orbit}
   \end{figure*}
%
%
   \begin{figure*}
   \centering
   \includegraphics[width=21cm]{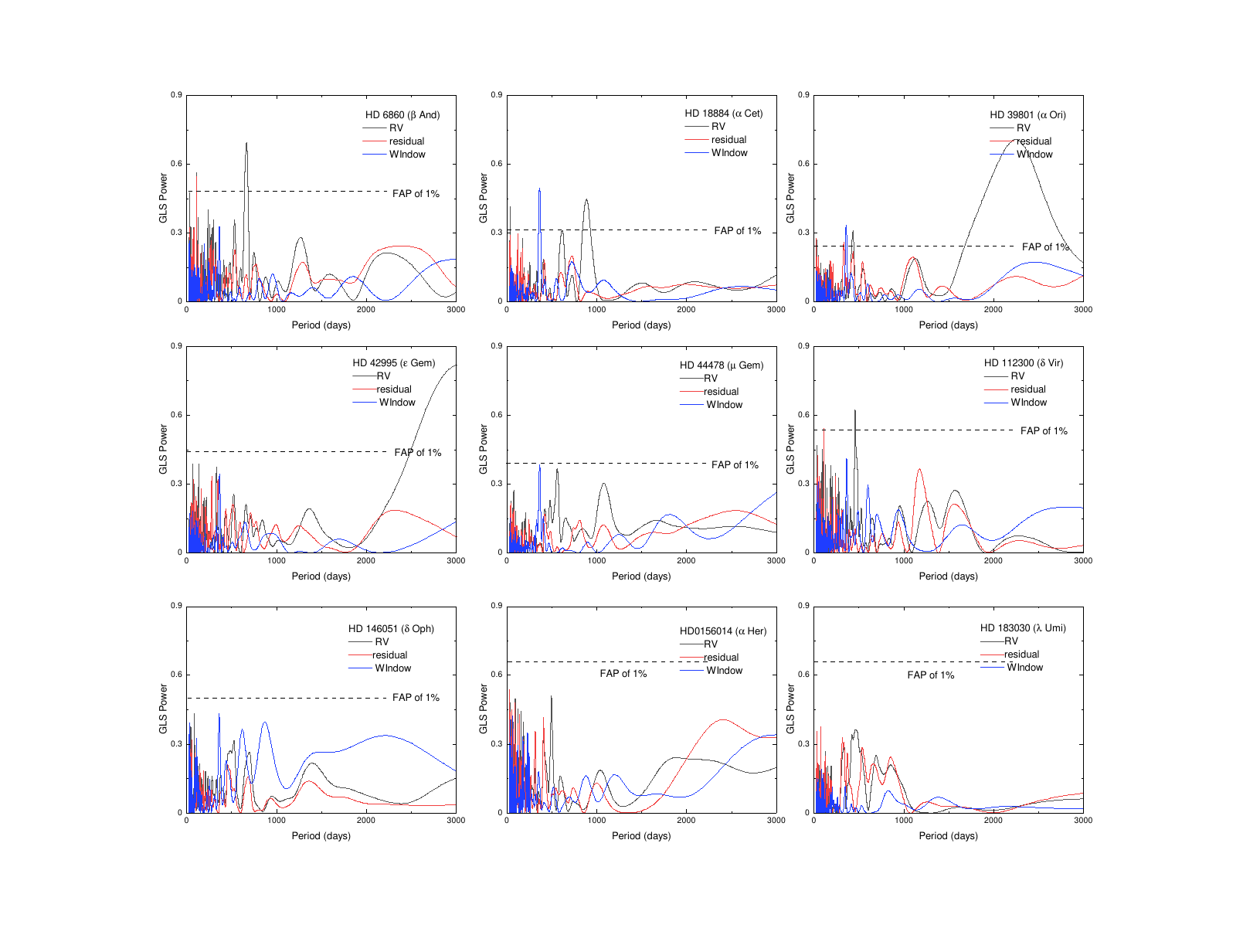}
      \caption{GLS power for nine M giants. The horizontal lines in each panel correspond to a 1\% FAP.
              }
         \label{f4_all_power}
   \end{figure*}

\subsection{Fundamental Parameters}
The basic stellar parameters were based on the \textit{HIPPARCOS} catalog \citep{1997yCat.1239....0E} and Gaia DR2 \citep{2018A&A...616A...1G}.
The stellar characteristics of M giants are not well known.
Due to the disagreement in the mass of the host star as measured by different authors and by different methods, the planet mass is subject to some uncertainty.
In order to derive the stellar mass, one needs the stellar parameters, particularly the effective temperature and gravity.
All stellar masses in this work are from the literature because the mass estimates are consistent with each other within acceptable error ranges for a given star.
Figure~\ref{f2_HR} shows the position of target stars in the H-R diagram. The sizes of the points are proportional to the stellar radii. Our RV measurements for the nine M giants are listed in Appendix A.

\subsection{\textbf{Rotational Velocity}}
We estimate the stellar rotational velocity using the line-broadening model devised by \citet{2008PASJ...60..781T}.
The observed stellar spectra were fitted by convolving the intrinsic spectrum model and the total macro-broadening function.
To determine the degree of line broadening, we used the automatic spectrum-fitting technique by \citet{1995PASJ...47..287T} when the spectrum was within the wavelength range of 6080 to 6089 {\AA}. This technique employs seven free parameters that specify the best fitting solution in relation to the abundances of six elements (Si, Ti, V, Fe, Co, and Ni).
We used the \citet{1989ApJ...347.1021G} assumption that the macro-turbulence depends only on the surface gravity.
In this work, we measured the macro-broadening velocity $v_{M}$ and estimated the projected rotational velocity $v_{\mathrm{rot}}$ sin($i$) for all stars.
In the literature, the $v_{\mathrm{rot}}$ sin($i$) values are from various sources obtained by different methods.
\citet{2008AJ....135..209M} used the cross-correlation of the observed spectra against templates drawn from a library of synthetic spectra calculated by \citet{1992IAUS..149..225K} for different stellar atmospheres.

We calculated the rotational velocity for nine targets. However, the line-broadening model we adopt does not apply when the surface temperature is lower than $\sim$ 3,500 K (HD~39801, HD~156014, and HD~42995) or for stars with very weak surface gravity (HD~39801 and HD~156014), and we were able to determine the rotational velocities for only three targets.

Based on the rotational velocities and the stellar radius, we derived the ranges of the upper
limits for the rotational periods of 7,900 days for HD~6860, 6250 days for HD~18884, and 4200 days for HD~112300.
The stellar parameters are listed  in Table.~\ref{tab1}.


\section{Orbital solutions and data analysis}

We used RVI2CELL \citep{2007PKAS...22...75H} for the best fit to determine the orbital parameters. 
In order to determine the periodicity in the BOES RV variations, we used the generalized Lomb-Scargle periodogram (GLS; \citealt*{2009A&A...496..577Z}). The GLS provides more accurate frequencies, is less susceptible to aliasing, and confers a much better determination of the spectral intensity.
All RV measurements are shown in Figure~\ref{f3_all_orbit} and the GLS outcomes for all objects are shown in Figure~\ref{f4_all_power}.
The GLS false-alarm probability (FAP) was calculated using a bootstrap randomization process \citep{2003A&A...403.1077K}. The RV data were shuffled 200,000 times keeping the times fixed. Periodogram was used to check sinusoidal periodicity in data set for first interpretation of the data. However, to assess error bars on the parameters and  to get a more robust fit of the data, we used the Markov Chain Monte Carlo (MCMC) algorithm and Exo-Striker \citep{2019ascl.soft06004T} to perform all analysis.

The long-period RV variations in giants are, in general, related to stellar rotation, pulsation, and/or reflex orbital motion  by a companion.
To uncover the origin of the detected RV variation of our targets, we studied the photometric variations, bisector variations, and stellar activities.

First, the variations in \textit{HIPPARCOS} photometric data were analyzed to determine the cause of stars showing RV periodicity.
Second, stellar rotational modulations by inhomogeneous surface features can create variable asymmetries in spectral line profiles \citep{2001A&A...379..279Q}.
Two bisector quantities were calculated from the line profile at two flux levels (40\% and 80\%) (the bisector velocity span BVS $\equiv$ [V$_{top}$ -- V$_{bottom}$] and the bisector velocity curvature BVC $\equiv$ [V$_{top}$ -- V$_{center}$] -- [V$_{center}$ -- V$_{bottom}$].

Finally, we used four kinds of indicators to check chromospheric  activities. 
The EW variations of Ca II H \& K, Balmer lines (H$_{\alpha}$, H$_{\beta}$), sodium lines, and Ca II 8662 {\AA} lines are frequently used as chromospheric activity indicators because they are sensitive to stellar atmospheric activity.
Such activity could have strong effects on the RV variations.
The efficiency of the short wavelength region (Ca II H \& K) of the BOES spectra is low. Moreover, another activity indicator, the Ca II 8662~{\AA} line, is also not suitable because significant fringing and saturation of our CCD spectra were found at wavelengths longer than 7500~{\AA}.
While H$_{\alpha}$ is sensitive to atmospheric stellar activity \citep{2003A&A...403.1077K} and useful for measuring
variations, it is not easy to estimate the H line EW due to line blending in the stellar spectra and telluric
lines if any exist.
To avoid nearby blending lines (i.e. Ti~I 6561.3, Na~II 6563.9, and ATM $\rm H_{2}O$ 6564.0 $\rm\AA$), we measured the H line EWs using a band pass of $\pm$ 1.0 $\rm\AA$ centered on the core of the H lines.
In the EW calculation of the sodium lines, it is important to avoid nearby blending lines and weak telluric lines.
We measured the sodium D lines at 5889.951 {\AA} and 5895.924 {\AA} using a band pass value of $\pm$ 0.5 $\rm\AA$ centered at the core of the sodium lines.

\section{Results\label{sec:nature}}

\begin{figure}[t]
\centering
\includegraphics[width=10cm]{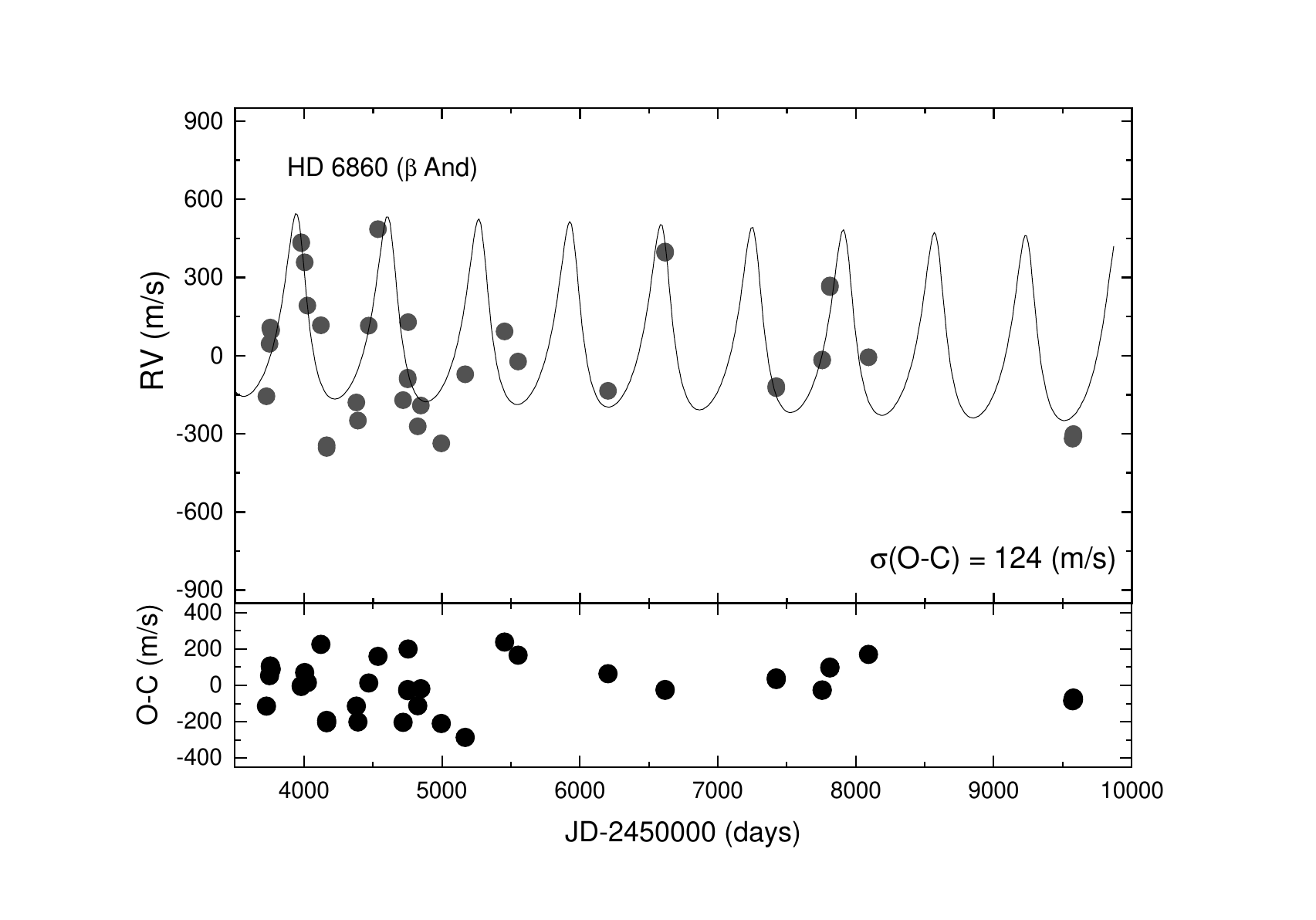}
\caption{RV measurements (top panel) and the residuals (bottom panel) for HD~6860 from December 2005 to December 2021.
The solid line is the orbital solution with a period of 664 days.  \label{orbit1}}
\end{figure}

\begin{figure}[t]
\centering
\includegraphics[width=10cm]{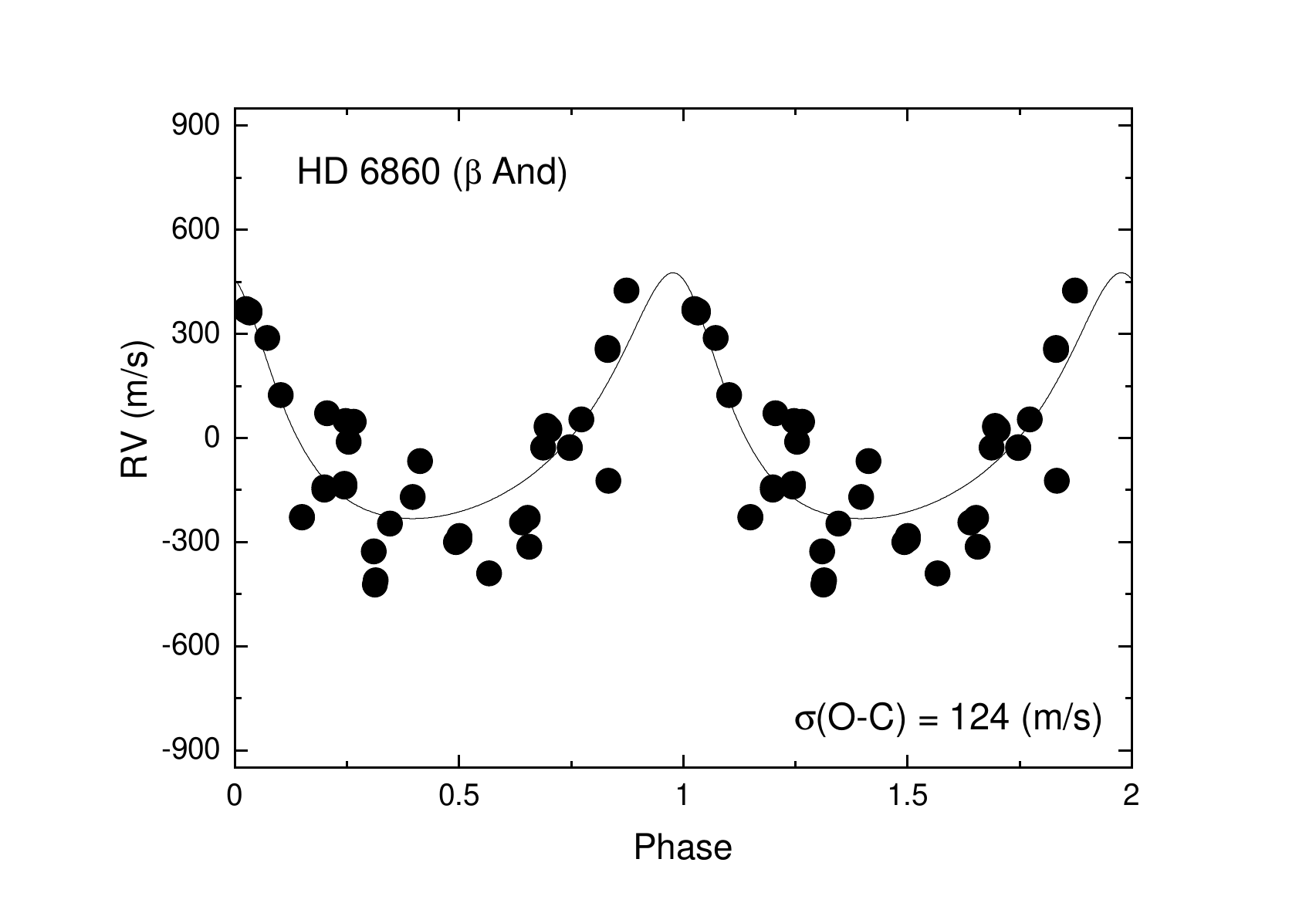}
\caption{Phase diagram for HD 6860. \label{phase1}}
\end{figure}

\begin{figure}
\centering
\includegraphics[width=9.5cm]{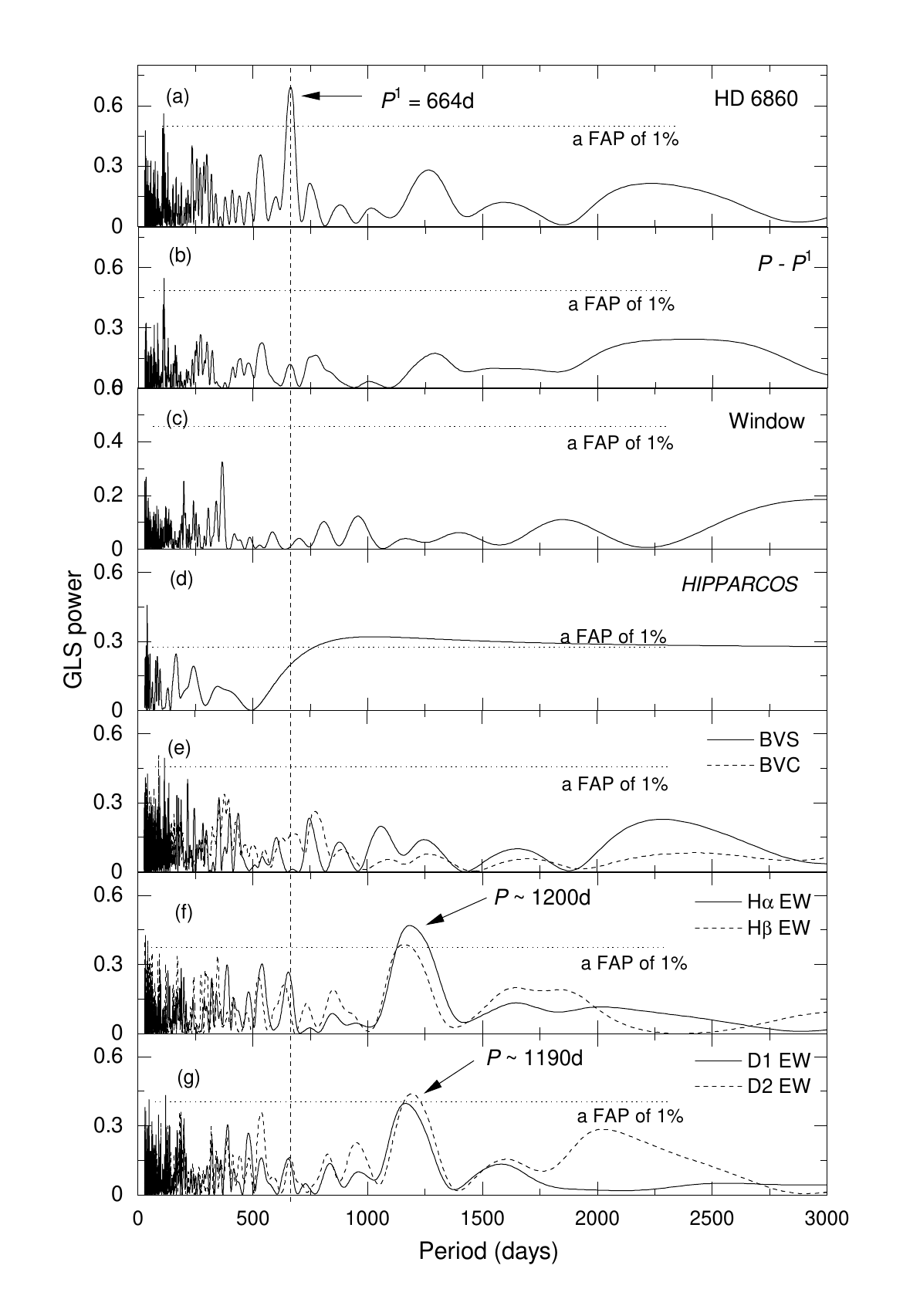}
\caption{(a) GLS periodogram for RVs of \mbox{HD 6860}, (b) The same periodogram for the residual RVs after subtracting the strong peak of 664d. (c) Periodogram of our time sampling (window function), (d) Periodogram of the \textit{HIPPARCOS} photometric data, (e) Periodograms of the line bisector span (BVS) and the line bisector curvature (BVC), (f) H EW variations, and (g) Sodium line EW variations. The vertical dashed lines indicate orbital periods of 661 days. The horizontal lines in each panel correspond to a 1\% FAP.
\label{power1}}
\end{figure}

\subsection{HD 6860 ($\beta$ And)\label{sec:nature}}
HD 6860 ($\beta$ And) is a bright star in the northern constellation of Andromeda and has an average apparent visual magnitude of 2.06. It is approximately at 60 parsecs from the sun according to its Gaia parallax  \citep{2018A&A...616A...1G}.
The total extinction towards this star in the visual band is about 0.06 mag.
This star has a small RV of 0.7 km s$^{-1}$ compared to the large tangential velocities from the proper motion values.
HD~6860 is known to be a single red giant of spectral type M0 III. It is suspected of being a semiregular variable star whose apparent visual magnitude varies by a magnitude of 0.014.

We obtained 44 spectra in total for \mbox{HD 6860} from December 2005 to December 2021.
The keplerian fit to RV data yields the following orbital parameters of the \mbox{HD 6860} system: an orbital period of 663.87$^{+4.61}_{-4.31}$ days, a semi-amplitude of 373.97 $^{ +22.63}_{-27.86}$ m s$^{-1}$, and an eccentricity of 0.28$^{+0.10}_{-0.09}$.
Figure~\ref{orbit1} shows RV variations  for HD~6860. The RV measurements phased to the orbital period are shown in Fig.~\ref{phase1}.
Assuming 2.49 $M_{\odot}$ \citep{2011A&A...533A.107D} for the mass of HD~6860, the companion has a  minimum mass of 28.26$^{+2.06}_{-2.18}$ $M_{J}$.
In order to ensure the reliability of our periodogram analysis, we compute the FAP.
Regarding the prominent peak at 661 days  in the RV power spectrum, we computed the preliminary FAP, finding it to be lower than 1 $\times 10^{-6}$\% level on the periodogram of the RVs.
Figure~\ref{f14_hd6860_mcmc} in the Appendix B shows the posterior distribution and correlations between all parameters sampled with our MCMC.

The \textit{HIPPARCOS} photometric data contain 59 observations for \mbox{HD 6860}. The data  has RMS scatter less than the magnitude of 0.014.
The GLS periodogram of the \textit{HIPPARCOS} measurements  shows a significant period of 41 days.
However, the short apparent periodicity of 41 days for HD 6860 is too short to originate from an orbiting planet as the semi-major axis would then be inside the stellar photosphere.
To search for variations in the spectral line shape, we selected the V1 6039.722 line, which is an unblended spectral feature with a high flux level located beyond the I$_{2}$ absorption region, meaning that contamination should not affect our bisector measurements.
We carried out GLS period analyses of the bisectors and no significant peaks were found as shown in panel (e) of Fig.~\ref{power1}.

The RMS values of the H$_{\alpha}$ and H$_{\beta}$ EWs indicate less than 0.4\% variation.
The GLS periodogram of the H line EW variations for \mbox{HD~6860} are shown in panel (f) of Fig.~\ref{power1}.
Significant peaks at 1,200 days similar to those of the H$_{\alpha}$ and H$_{\beta}$ lines were found and judged to be unrelated to the RV period.
The RMS values in the sodium line D1 EWs and D2 EWs shows less than 0.1\% variation in both stars.
The GLS periodograms of the sodium D EW variations are shown in Fig.~\ref{power1} (g) for \mbox{HD 6860}.
The sodium lines of \mbox{HD 6860} show important periods at around 1,190 days, very similar to those shown in the H lines (1,200-days).
Therefore, the 1,200-day period is quite likely to be related to chromospheric activity.

\begin{figure}[!t]
\centering
\includegraphics[width=10cm]{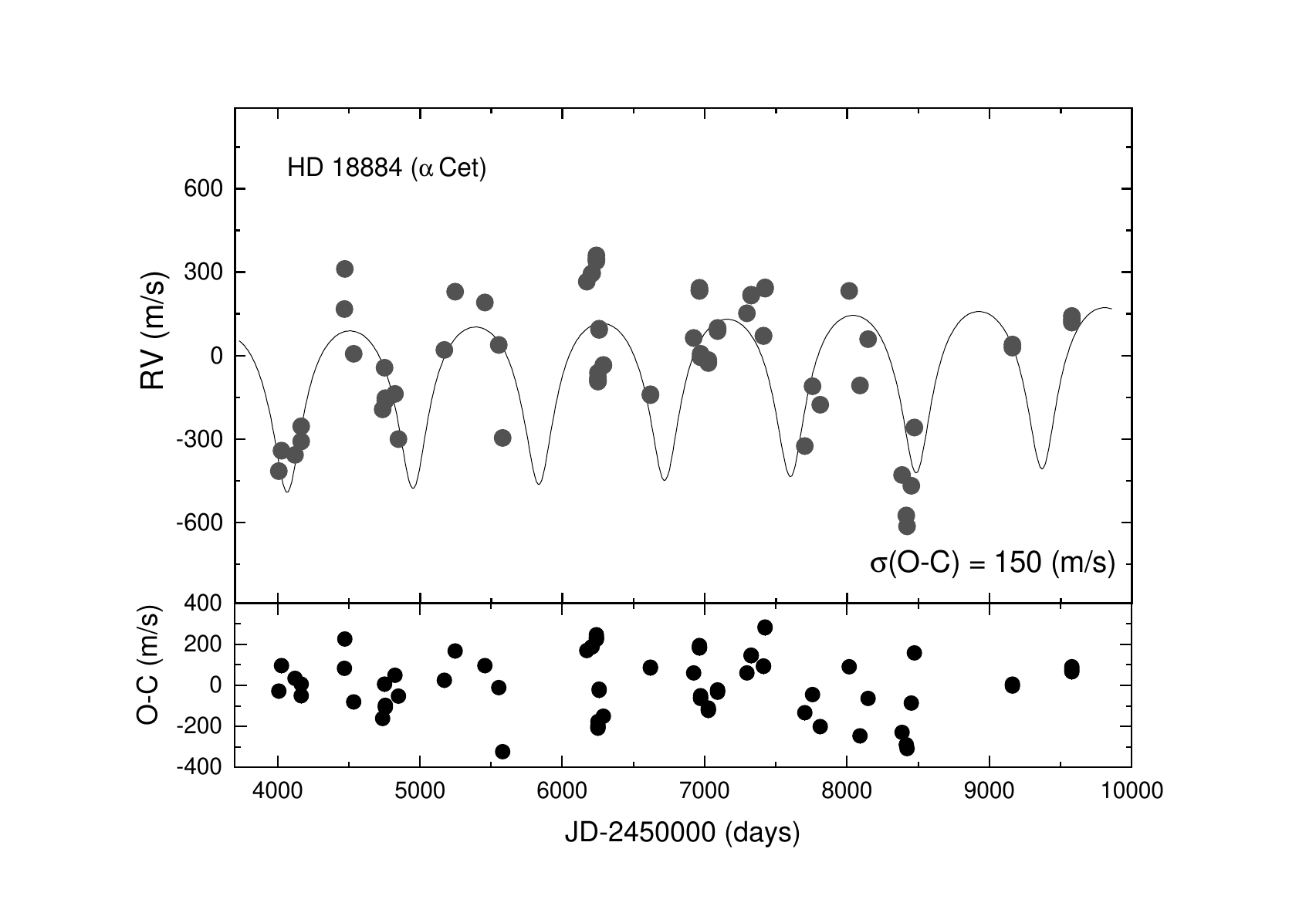}
\caption{RV measurements (top panel) and the residuals (bottom panel) for HD~18884 from December 2005 to December 2021.
The solid line is the orbital solution with a period of 885 days.  \label{orbit2}}
\end{figure}

\begin{figure}[!t]
\centering
\includegraphics[width=10cm]{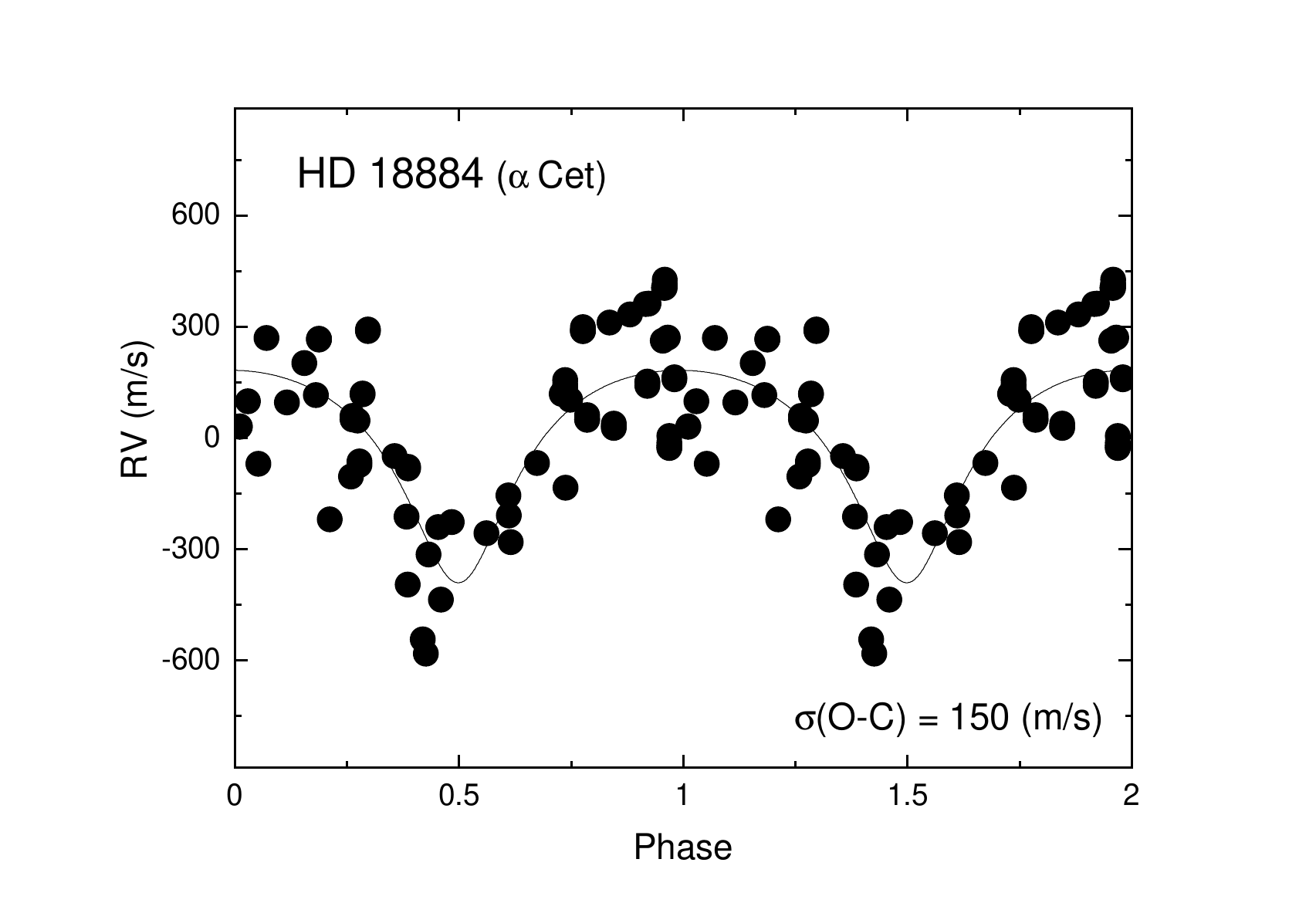}
\caption{Phase diagram for HD 18884. \label{phase2}}
\end{figure}

\begin{figure}[!h]
\centering
\includegraphics[width=10cm]{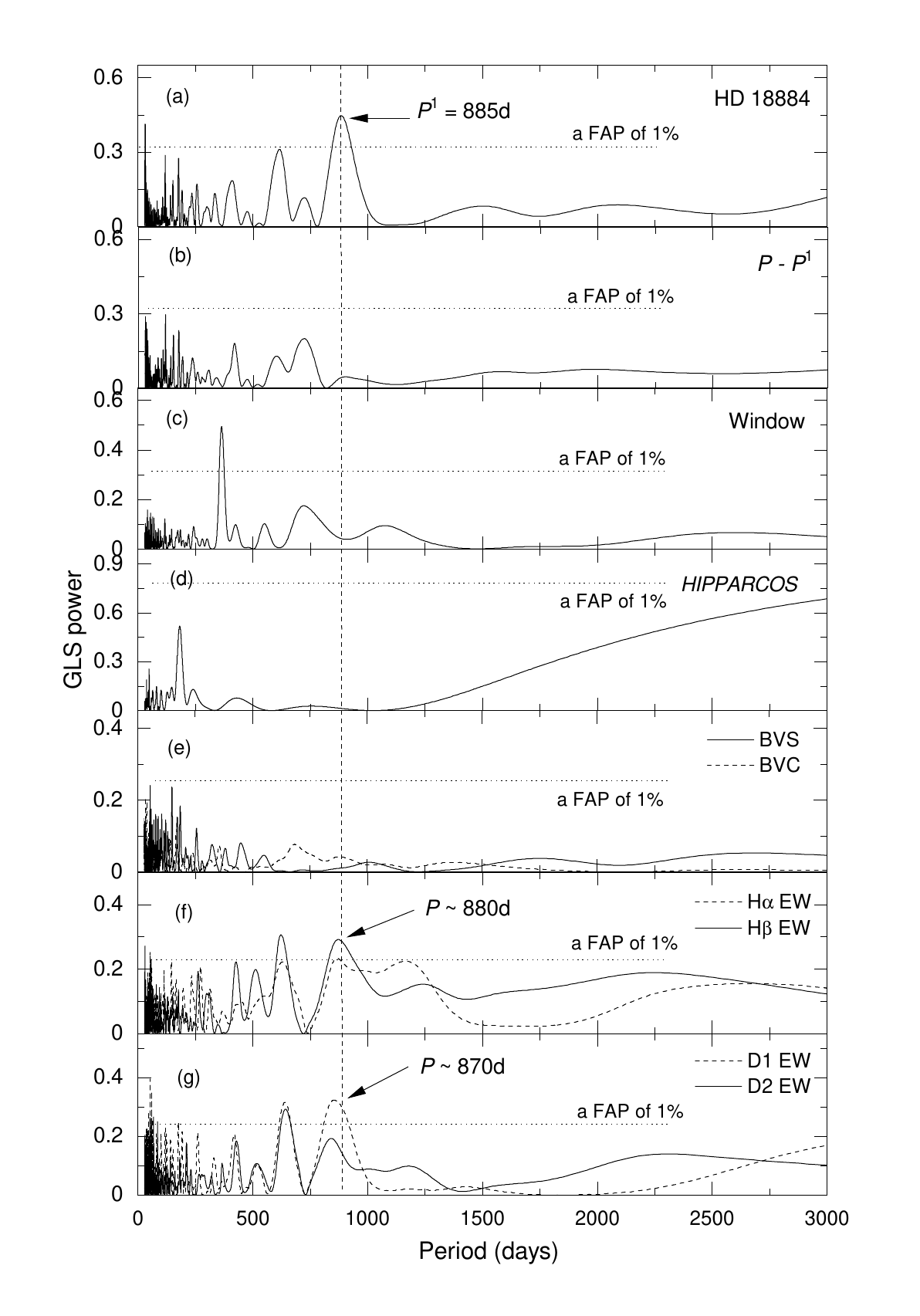}
\caption{(a) GLS periodogram for RVs of \mbox{HD 18884}, (b) The same periodogram for the residual RVs after subtracting the strong peak of 885d. (c) Periodogram of our time sampling (window function), (d) Periodogram of the \textit{HIPPARCOS} photometric data, (e) Periodograms of the line bisector span (BVS) and the line bisector curvature (BVC), (f) H EW variations, and (g) Sodium line EW variations. The vertical dashed lines indicate orbital periods of 661 days. The horizontal lines in each panel correspond to a 1\% FAP.
\label{power2}}
\end{figure}

\subsection{HD 18884 ($\alpha$ Cet)\label{sec:nature}}

HD 18884 is a cool luminous red giant about 77 parsecs away from the Sun and known to be an LPV.
It is at the AGB evolutionary stage, during which hydrogen and helium are exhausted at the core  \citep{1992AJ....104..275E} and will most likely become a highly unstable star similar to Mira, before finally shedding its outer layers and forming a planetary nebula, leaving a relatively large white dwarf remnant.

Over the 15-year period from October 2006 to December 2021, a total of 75 spectra for HD 18884 were collected.
The RV periodic variations and a phase diagram of HD~18884 are shown in Figs.~\ref{orbit2} and~\ref{phase2}.
The GLS periodogram of the RV measurements shows a significant peak at 885 days.
The orbital parameters of the system were derived as a semi-amplitude of 287 $\pm$ 53 m s$^{-1}$ and an eccentricity of 0.4 $\pm$ 0.2.
A relatively large residual (RMS of 150 m s$^{-1}$) corresponding to the characteristics of late-type stars was also noted, though no secondary significant peak was found.
If we adopt a mass of 2.3 $M_{\odot}$ for HD~18884 \citep{2006A&A...460..855W}, the best-fit Keplerian model yields a minimum mass of 21 $M_{J}$  for its companion.

The 57 observations of \textit{HIPPARCOS} photometric data were used for \mbox{HD 18884}, which shows a RMS scatter of less than a magnitude of 0.014. No significant peak was found from the GLS periodogram.
To search for variations in the spectral line shape, we selected the V1 6039.722 {\AA} line. GLS period analyses show no significant peaks (Figure~\ref{power2}).

The RMSs in the H$_{\alpha}$ and H$_{\beta}$ EWs indicate less than 0.4\% variation.
The GLS periodogram of the H line EW variations for \mbox{HD 18884} are shown in panel (f) of Fig.~\ref{power2}. Significant peaks at 880 days similar to those of the RV variation were found.
The RMS value in the sodium line D1 EWs and D2 EWs shows less than 0.1\% variation in both stars.
The GLS periodograms of the sodium D EW variations are shown in Fig.~\ref{power2} (g).
Given that the 870-day period found in HD~18884 nearly coincides with the H lines and RV period, we reckon that the main cause of the RV period is stellar atmospheric activities.

\subsection{HD 39801($\alpha$ Ori)\label{sec:nature}}

The M2 Iab supergiant HD 39801 ($\alpha$ Ori) is an ideal laboratory to search for exoplanets around massive stars and to study their properties.
The star is classified as a semiregular variable with an SRC sub-classification with a period of 2,335 days (6.39 yr) \citep{2009yCat....102025S}.
 The absolute luminosities and photospheric radii are now sufficiently well determined to warrant a new investigation of the constraints on models for this star.
The list in the Catalog of Components of Double \& Multiple stars (CCDM) contains at least four combined celestial bodies, all of which are within three arcmins of the star. However, little is known about these except for their position angles and apparent magnitudes \citep{2002yCat.1274....0D}.
Thus, HD 39801 has no known orbital companions and, therefore, its mass has not yet been determined with any certainty.
A pulsating semiregular variable star such as HD 39801 is subject to multiple cycles of increasing and decreasing brightness due to changes in its size and temperature.
This star is usually considered to be a runaway star without any companion \citep{2013EAS....60..307V}. The presence of companions has been suggested from spectroscopic and polarimetric observations \citep{1986ApJ...308..260K}.

Over the 13 year period from January 2008 to April 2021, we collected 92 spectra for HD~39801.
The resulting power spectrum displays a broad, highly significant peak centered at  a period of 2,245 days (Figure~\ref{f4_all_power}).
This period is very similar to 2,335 days presented by \citet{2009yCat....102025S}.
Assuming the stellar mass of 16.5$-$19 $M_{\odot}$, HD~39801 has approximately one solar mass companion at a distance of 8.5$-$9.0 AU. This result is similar to that in earlier work \citep{2009yCat....102025S}.

From November  2019 to March  2020, HD~39801 experienced a historic dimming of visible brightness. The apparent brightness  has decreased to about 1.6 magnitudes and the southern hemisphere of HD~39801 was ten times darker than the northern hemisphere in the visible spectrum during its great dimming in February 2020 \citep{2021Natur.594..365M}.
Unfortunately, we could not confirm the great dimming because it was not observed in February  2020 using the BOES.

\subsection{HD 42995 ($\eta$ Gem)\label{sec:nature}}

HD 42995, an RGB, is a triple system composed of a primary M-type giant and a close companion, as spectroscopically confirmed.
This star has been classified as a semi-regular variable with SRa-type variability.
This variability closely approximates Mira variables, but the amplitude is lower.  Many long-period variables show long secondary periods with typically ten times longer than the main period with an amplitude up to one magnitude at visual wavelengths. However, such changes have not been detected for HD 42995.
The main period has been known as an average of 234 days \citep{2008JAVSO..36..139P}, which is very close to 235 days period from \textit{HIPPARCOS} photometric data.
Little is known about the companion, although HD~42995 is a sixth magnitude bright star. It is assigned to be a G0 spectral type and giant on the basis of its brightness \citep{1998A&A...330..225H}.
The orbit calculated in 1944 is essentially unchanged today with a period of 2,983 days and an eccentricity of 0.53. Observers have found signs of an eclipse corresponding to the derived orbit. However, the evidence was considered inconclusive \citep{1944ApJ...100...63M}.
On the other hand, the companion is suspected to be an M-type star given the appearance of its spectrum \citep{1998A&A...330..225H}.

We obtained 48 spectra from October 2006 to April 2021.
The power spectrum displays a broad, highly significant peak centered at a period of 3,001 $\pm$ 11 days and an eccentricity of 0.53 $\pm$ 0.05 (Figure~\ref{f4_all_power}),  very similar to the 2,983 days and 0.53 found by \citet{1944ApJ...100...63M}.
Assuming a mass of 2.5$M_{\odot}$ \citet{1998A&A...330..225H}, HD 42995 has a companion of  1.2 $M_{\odot}$ (minimum mass) at a distance of 5.5 AU.

\subsection{HD 44478 ($\mu$ Gem)\label{sec:nature}}

HD 44478 is a slow irregular type variable in the AGB stage of spectral type M3 III.
Its brightness in the V-band varies between magnitudes +2.75 and +3.02 over a 27 day period, along with a 2,000 day period of long-term variation \citep{2001PASP..113..983P}.

We determined the orbital period of 560 days from the RV measurement for HD 44478 over 15 years (see Figure 4). A FAP is estimated to exceed 5\%. This usually means that it is unlikely to be a statistically significant signal. There was no evidence for a 27 day period nor a long-term variation of 2000 days. However, the periodogram  for \textit{HIPPARCOS} photometric data shows 27 day period with a brightness change of 0.04, the same as that reported by \cite{2001PASP..113..983P}. It suggest that it is probably due to low-order radial pulsation \citep{2001PASP..113..983P} and  a long-term variation  may be due to rotation, or convection-induced oscillatory thermal modes in red giants known as a new type of stellar oscillation proposed by \cite{2000ASPC..203..379W}.

\begin{figure}[t]
\centering
\includegraphics[width=10cm]{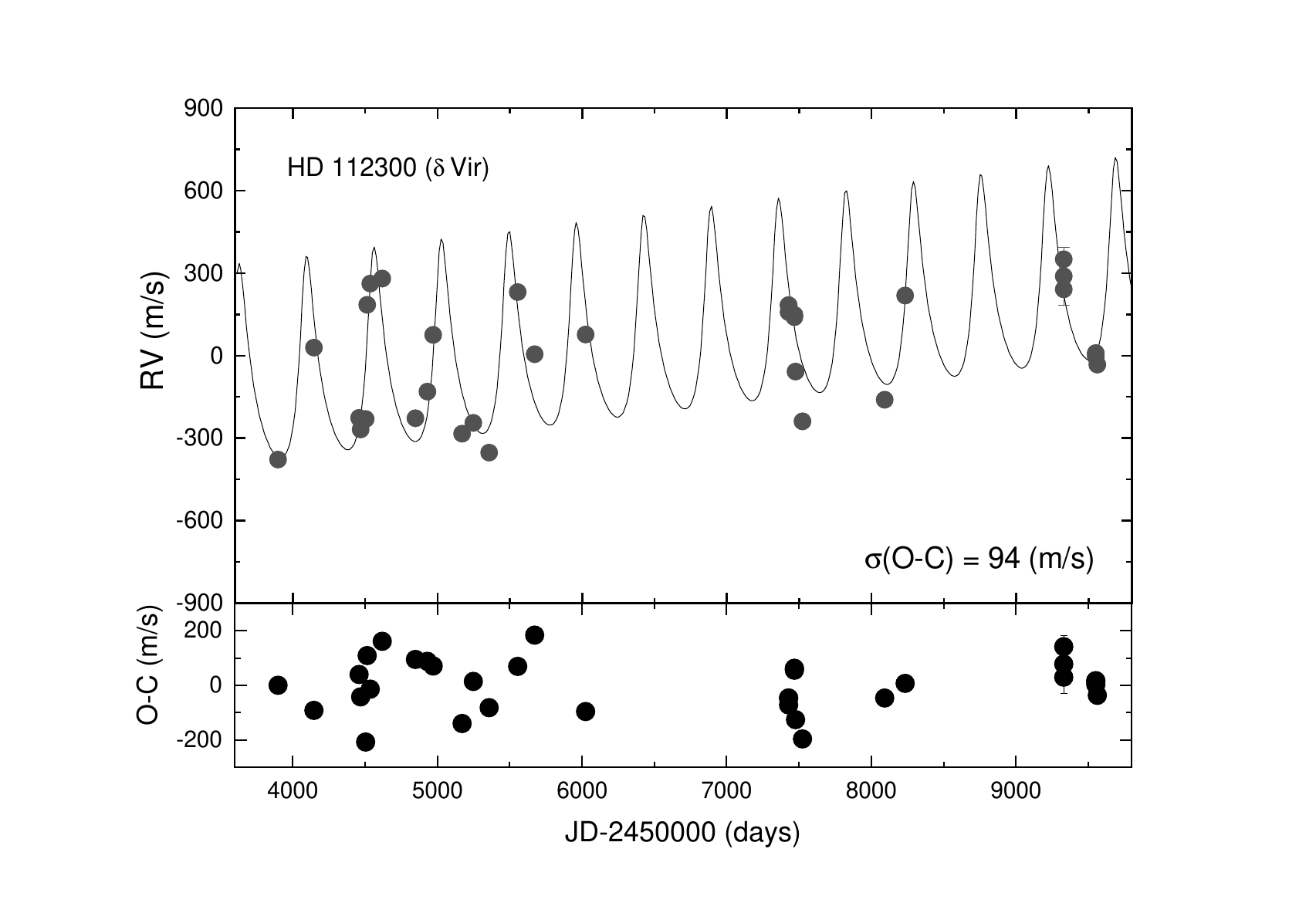}
\caption{RV measurements (top panel) and the residuals (bottom panel) for HD~112300 from June 2006 to December 2021.
The solid line is the orbital solution with a period of 467 days.  \label{orbit3}}
\end{figure}

\begin{figure}[t]
\centering
\includegraphics[width=10cm]{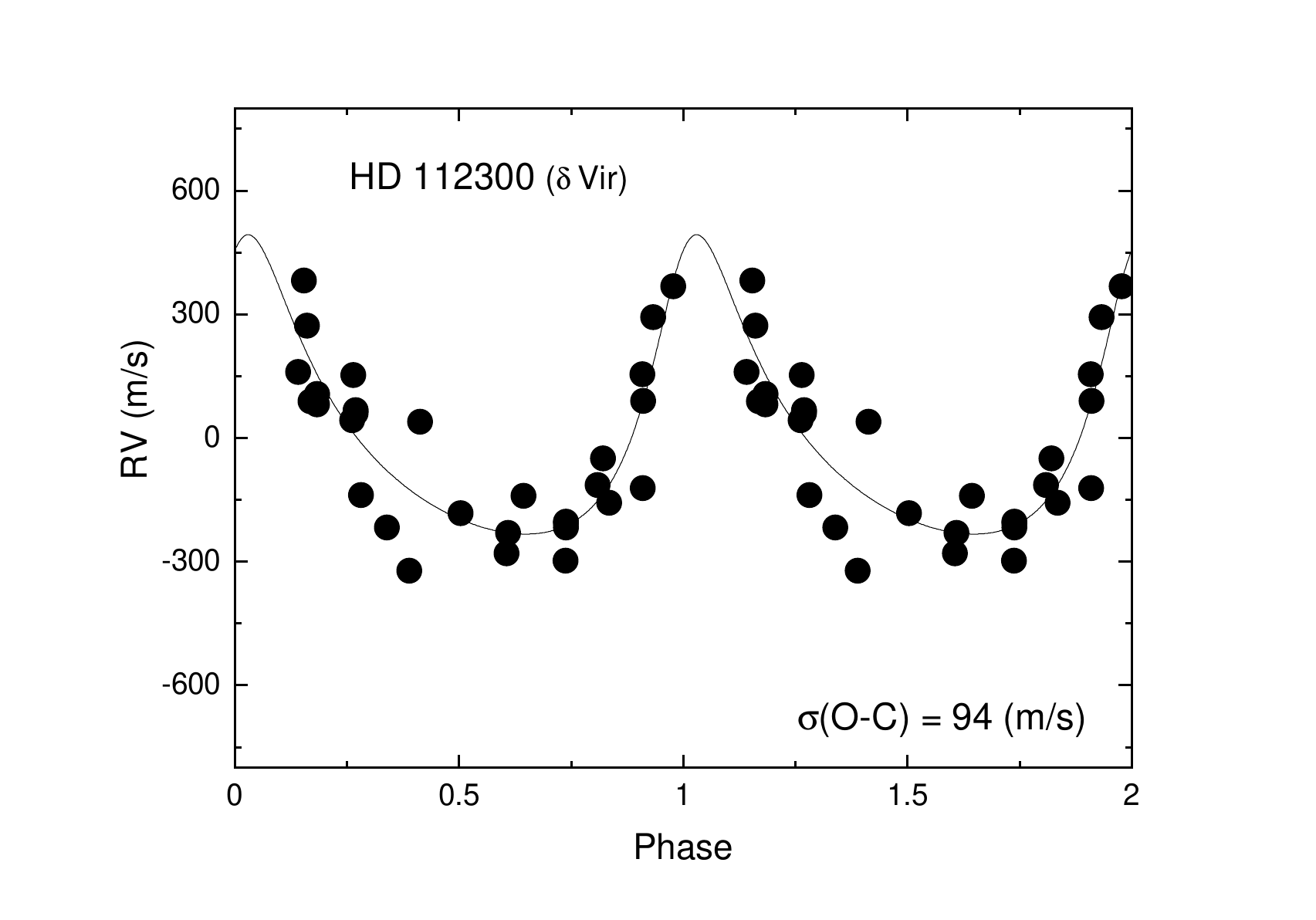}
\caption{Phase diagram for HD112300. \label{phase3}}
\end{figure}

\begin{figure}[h!]
\centering
\includegraphics[width=10cm]{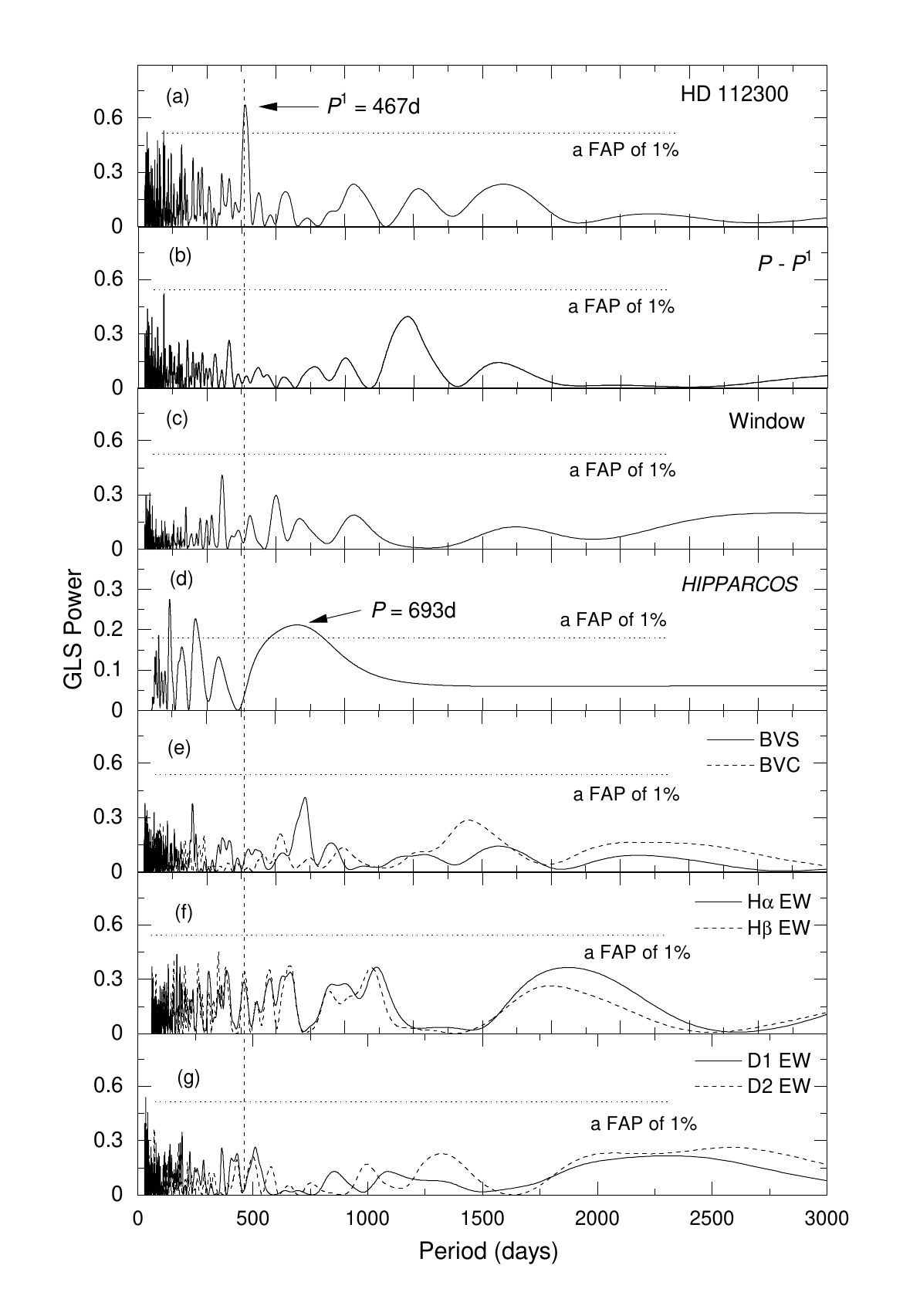}
\caption{(a) GLS periodogram for RVs of \mbox{HD 112300}, (b) The same periodogram for the residual RVs after subtracting the strong peak of 467 days. (c) Periodogram of our time sampling (window function), (d) Periodogram of the \textit{HIPPARCOS} photometric data, (e) Periodograms of the line bisector span (BVS) and the line bisector curvature (BVC), (f) H EW variations, and (g) Sodium line EW variations. The vertical dashed lines indicate orbital periods of 467d. The horizontal lines in each panel correspond to a 1\% FAP.
\label{power3}}
\end{figure}

\subsection{HD 112300 ($\delta$ Vir)\label{sec:nature}}

HD 112300 is an M1 III star with a mass of 1.4 solar masses and  effective temperature of  $\sim$ 3,660 K \citep{2017MNRAS.471..770M}.
This star exhibits semi-regular variability with a very small amplitude showing a magnitude of approximately 0.04 according to \textit{HIPPARCOS} photometric data.
A frequency analysis of the observed optical curve shows pulsations of several cycles. Known periods found by \cite{2009MNRAS.400.1945T} are 13.0 days, 17.2 days, 25.6 days, 110.1 days, and 125.8 days.
HD 112300 may be a  binary star with a K-type dwarf companion and may have an orbital period exceeding of 200,000 years, but this has not been confirmed.

For HD~112300, 35 spectra in total were collected from June 2006 to December 2021.
The best-fit  model yielded a period of 466.63 $^{+1.47}_{-1.28}$ days, a semi-amplitude of 353.99$^{+13.63}_{-23.04}$ m s$^{-1}$, and eccentricity of $e$ = 0.36 $^{+0.06}_{-0.11}$. 
The RV also clearly shows a linear trend. Thus,  we included the linear slope as an unknown parameter in the orbital solution.
Assuming a stellar mass of 1.4 $\pm$ 0.3~$M_{\odot}$, we derived a minimum mass of a planetary companion $m$~sin($i$) = 15.83$^{+2.34}_{-2.74}$ $M_{\rm{Jup}}$ at a distance $a$ = 1.33 $^{+0.09}_{-0.11}$ AU from the host. 
Figure~\ref{orbit3} shows the periodic variations  for HD~112300.
and the RV phase diagram of the orbit is shown in Fig.~\ref{phase3}.
The posterior distribution and correlations between all parameters for HD~112300 sampled with our MCMC is shown in Appendix~\ref{f15_hd112300_mcmc}.

The \textit{HIPPARCOS} photometric data contain 93  observations for \mbox{HD 112300}.
They reveal considerable scatter with a magnitude of 0.022. GLS periodogram analyses indicate strong peaks at 138 d, 251 d, and 693 d for \mbox{HD 112300}, which are inconsistent with the RV period of 467 days.
To search for variations in the spectral line shapes,  Ba II 6141.713 {\AA} line was used. 
The GLS analysis of the Barium time series produces no significant signals [panel (e) of Fig.~\ref{power3}].

The RMS in the H$_{\alpha}$ and H$_{\beta}$ EWs indicate less than 0.4\% variation.
The GLS periodogram of the H line EW variations for \mbox{HD 112300} are shown in panel (f) of Fig.~\ref{power3}.
There is no significant peak.
The RMS value in the sodium line D1 EWs and D2 EWs shows less than 0.1\% variation in both stars.
The GLS periodograms of the sodium D EW variations are shown in Fig.~\ref{power3}(g).
The significant peak is below 30 days for \mbox{HD 112300}. 
This short period has nothing to do with the existence of planets around giants, but is rather related to the stellar activity.

\subsection{HD 146051 ($\delta$ Oph)\label{sec:nature}}

The high-proper-motion star HD 146051 has a stellar classification of M0.5 III. It is currently on the AGB. The measured angular diameter of this star, after correction for limb darkening, is 10.47 $\pm$ 0.12 mas \citep{2005A&A...431..773R}. This yields a physical size of about 54 times the radius of the Sun.
The effective temperature of the outer atmosphere is relatively cool at $\sim$ 3,800 K \citep{2017MNRAS.471..770M}. It is listed as a suspected variable star that may change by magnitude of 0.03 in the visual band.

We obtained 39 spectra from June 2006 to March 2021. GLS results show no significant peak in the RV measurements (Figure~\ref{f4_all_power}).

\subsection{HD 156014 ($\alpha$ Her)\label{sec:nature}}

The AGB star HD 156014 (M5 Ib-II) is a LPV which is the primary star of a triple system.  The primary star forms a visual binary pair with a second star, which is itself a spectroscopic binary \citep{2013AJ....146..148M}. HD 156014 A and B are more than 500 AU apart, with an estimated orbital period of approximately 3,600 years. HD~156014 A is a relatively massive red bright giant; however, RV measurements suggest a companion with a period on the order of a decade (referrenced from the "Washington Double Star Catalog").
The full range of its brightness is from magnitude 2.7 to magnitude 4.0 \citep{2009yCat....102025S}. However, it usually varies within a much smaller range of around 0.6 magnitude \citep{2001PASP..113..983P}.
The two components of HD 156014B are a primary yellow giant star and a secondary, yellow-white dwarf star in a 51.578 day orbit.

We obtained 25 measurements of HD 156014 spanning more than nine years from June 2010 to October 2019. There is no significant power in the RV data and only a period of 497.6 days with a FAP of 5\%, which is not related  to the brightness variation.

\subsection{HD 183030 ($\lambda$ UMi)\label{sec:nature}}

HD 183030 is an AGB star showing semiregular variability.
\textit{HIPPARCOS} photometric data show that HD~183030 is a LPV star with a period of 647 days.
Its brightness varies by a magnitude of approximately 0.1 in the visual band.

Total of 30 spectra were taken from June 2010 to January 2022 using the BOES.
Precisely measured RV data reveal no significant signals (Figure~\ref{f4_all_power}).
Perhaps the data are insufficient to provide meaning with which to reach a reliable scientific conclusion regarding a LPV for HD~183030.

\begin{table}
\begin{center}
\caption{Radial pulsation modes for HD 6860 b and HD 112300 b.}
\label{tab2}
\begin{tabular}{lccc}
\hline
\hline
    Mode                      &               &  HD 6860    & HD 112300  \\
\hline
    Fundamental period       & [days]        & 44      &  39  \\
    Pulsation period         & [days]        & 14      &  15  \\
    Pulsation RV amplitude   & [m s$^{-1}$]  & 157     &  84  \\
\hline
\end{tabular}
\end{center}
\end{table}
%

\begin{table}[h!]
\renewcommand{\thetable}{\arabic{table}}
\centering
\caption{Orbital solutions for HD 6860 b and HD 112300 b probed by the MCMC simulation.} \label{tab3}
\begin{tabular}{lcc}
\hline
\hline
parameter	& HD 6860 b    & HD 112300 b \\
\hline
P (days)			& 663.87$^{+4.61}_{-4.31}$	& 466.63 $^{+1.47}_{-1.28}$ \\
K (m s$^{-1}$)		& 373.97 $^{ +22.63}_{-27.86}$	&  353.99$^{+13.63}_{-23.04}$ \\
$e$					& 0.28$^{+0.10}_{-0.09}$     &  0.36 $^{+0.06}_{-0.11}$ \\
$T_{periastron}$ (JD)& 2453272.89$^{+65.23}_{-65.13}$ &  2453611.56$^{+27.9}_{-27.9}$ \\
$\omega$ (deg)		& 7.54 $^{+10.72}_{-4.99}$ & 333.29$^{+9.74}_{-9.81}$\\
MA$_b$ (deg)		& 247.93 $^{+6.03}_{-10.13}$ & 221.80$^{+9.15}_{-9.86}$\\
$\lambda_b$ (deg)		& 256.17 $^{+6.10}_{-5.46}$ & 194.80$^{+6.89}_{-7.10}$\\
$m$ sin $i$ ($M_{J}$)& 28.26$^{+2.05}_{-2.17}$ & 15.83$^{+2.33}_{-2.74}$ \\
$a$ (AU)         	 & 2.03$^{+0.01}_{-0.01}$ & 1.33 $^{+0.08}_{-0.11}$\\
$Slope$  (m s$^{-1} d^{-1}$) & -0.01088$^{+0.01374}_{-0.01276}$ &  0.05642$^{+0.00907}_{-0.01163}$ \\
RV offset (m s$^{-1}$) & 64.19$^{+21.39}_{-34.41}$	& -139.50 $^{+24.39}_{-21.52}$ \\
RV jitter (m s$^{-1}$) & 131.05$^{+20.02}_{-10.26}$	& 99.01 $^{+13.67}_{-13.67}$ \\
$N_{obs}$                     & 44 & 35	 \\
$chi^2$ (m s$^{-1}$)			  & 17.17  & 30.08 \\
rms (m s$^{-1}$)			  & 123.7  & 93.6 \\
\hline
\end{tabular}
\end{table}

%
\begin{table}
\begin{center}
\caption{The RV measurements of \mbox{HD 6860} recorded before the development of high-precision RV measurements.}
\label{tab4}
\begin{tabular}{lr}
\hline
\hline
    RV                   &  References      \\
    (km s$^{-1}$)        &                  \\
\hline
    1.5             & Campbell et al. (1911) \\
    2.0            & Campbell(1913)      \\
    0.5             & Campbell (1928)   \\
    0.6               & Shajn \& Albitzky(1932) \\
    3.2             & Wilson \& Joy (1952)   \\
    0.3             & Wilson(1953)  \\
    0.3             & Duflot et al. (1995)       \\
    0.06             & Famaey et al.(2005)    \\
    0.13             & Massarotti et al.(2008) \\
\hline
\end{tabular}
\end{center}
\end{table}
%

%
\begin{table}
\begin{center}
\caption{The RV measurements of \mbox{HD 112300} recorded before the development of high-precision RV measurements.}
\label{tab5}
\begin{tabular}{lr}
\hline
\hline
    RV                   &  References      \\
    (km s$^{-1}$)        &                  \\
\hline

    --19.6             & Campbell(1913)       \\
    --17.4             & Adams(1915)          \\
    --17.1             & Lunt(1919)   \\
    --17.9             & Campbell(1928)   \\
    --20.1             & Harper(1933)   \\
    --17.8             & Wilson(1953)   \\
    --18               & Fluks et al.(1994)   \\
    --17.5             & Duflot et al. (1995)   \\
    --17.8             & Wielen et al.(1999)    \\
    --18.87            & Famaey et al.(2005)    \\
    --18.14            & Massarotti et al.(2008)  \\

\hline
\end{tabular}
\end{center}
\end{table}
%

\section{Discussion\label{sec:dis}}

We selected nine bright M giants in the RGB or AGB stage and observed them for more than a decade to find low-amplitude and long-period RV variations using high-resolution spectroscopy.
Generally, most giant stars have intrinsic RV variations, pulsations and/or surface activities. In order to determine the nature of the RV variabilities, we  undertook comprehensively all relevant analyses. Also, a better understanding of how RV modulations behave as stars evolve during the RGB or AGB stage can be gained by a detailed study of all possible stellar-induced signals.

In general, giants reveal pulsation periods of a few days in the form of radial pulsations.
We estimated the fundamental periods, i.e., radial pulsation periods, and the expected  amplitudes for \mbox{HD 6860} and for \mbox{HD 112300} using the relationships devised by \citet{1995A&A...293...87K} in Table~ \ref{tab2}. The periods are far too short to explain the RV variations we observed. However, the expected RV amplitudes from the stellar oscillations are consistent with the RMS scatter in the RV orbital fits. Thus, the significant RV scatter observed with regard to the orbital solution can easily be explained by these stellar oscillations. We note that both stars are expected to have the smallest RV amplitudes attributable to stellar oscillations, and indeed have the smallest amounts of observed RV scatter.
The best-fit orbital parameters of the planetary signals in the final MCMC model are listed in Table~\ref{tab3}.

As a result of the long-term RV observations of the nine M giants, two long-term stellar companions were verified for HD~39801 and for HD~ 42995. In addition, low-amplitude and long-period RV periods were found around three M giants. Among them, the 885 day variations discovered for HD~18884 are strongly suspected to be due to stellar chromospheric activity, and the remaining two M giants HD~6860 and HD~112300 are found to harbor sub-stellar companions.

The variable M giant \mbox{HD 6860} (CSV 100088, NSV 414) and the M giant \mbox{HD 112300} (NSV 6026)  have been observed for photometric and spectroscopic variations since the early twentieth century.
A history of RV determinations of both stars recorded before the development of precise RV techniques is listed in Tables~\ref{tab4} and ~\ref{tab5}. However, there are few observed RV results for \mbox{HD 6860} and \mbox{HD 112300} despite the long interval. It has been found that the RV values vary with a range of $\sim$ 1.07 km s$^{-1}$ for \mbox{HD 6860} and $\sim$ 0.94 km s$^{-1}$ for \mbox{HD 112300}, which indicates the possibility of periodic RV variations.

The M giant HD 6860 reveals different long-term variations in its spectroscopic and chromospheric activity measurements.
Four types of chromospheric activity indicators were used to verify the cause of the RV origin. Interestingly, all indications show a nearly identical period of 1,190 -- 1,200 days, unrelated to the RV period of 661 d. This can be interpreted as a strong evidence that the cause of the RV variation at 1,200 days is chromospheric activity.

The M giant HD 112300 also shows different long-term variations in the spectroscopic and photometric measurements.
HD~112300 was discovered to be a variable with long-period of $\sim$ 693 days according to \emph{HIPPARCOS} measurements.
These variations are similar to those of a semiregular variable among late-type supergiants (SRC) or a long secondary-period variable (LSPV). SRCs are M supergiants that show semiregular, multi-periodic photometric variations ranging from several dozens to several thousands of days (i.e., BC Cyg, $\alpha$ Ori, $\mu$ Cep).
However, unlike one magnitude variation in the V band usually observed in a SRC variable, HD~112300 only shows a variation magnitude of 0.22 in the photometric measurements.

On the other hand, \citet{2006MNRAS.372.1721K} showed that $\sim$ 25\% of pulsating red supergiants show periodic brightness changes characterized by two distinct time scales: a few hundred days (first-period mode) and more than approximately 1,000 days (secondary-period mode), known as LSPVs.
Our analyses show that for HD 6680, the first period is caused by a sub-stellar companion and the long secondary period can be attributed to chromospheric activity. The first period for HD~112300 is also related to a sub-stellar companion and the long secondary period may stem from the rotational modulation of surface activities.
Further RV observations of these pulsating red supergiants will show if such distinct time scales generally have different origins as in HD~6860 and HD~112300.

As of January of 2022, approximately one hundred giant planets have been discovered around (super)giant stars. These giant planets are more massive than the giant planets found around solar-type stars. This may arise because giant stars are more massive than the Sun and more massive stars are expected to have more massive planets. However, the masses of the planets that have been found around giant stars do not correlate with the masses of the stars. Therefore, the planets could be growing in mass during the red giant stage of the hosting stars. This type of growth in planet mass could be partly due to accretion from stellar wind, although a much larger effect would be Roche lobe overflow causing a mass-transfer from the star to the planet when the giant expands out to the orbital distance of the planet \citep{2014A&A...566A.113J}.

To summarize,
we were observed nine bright M red giants since 2005 and found a significant periodic RV signals on HD~6860 and HD~112300 originated from sub-stellar companions.
The best Keplerian fit and MCMC simulation yields an orbital period of 663.87$^{+4.61}_{-4.31}$ days, a semi-amplitude outcome of 373.97 $^{ +22.63}_{-27.86}$ m s$^{-1}$, and eccentricity of 0.28$^{+0.10}_{-0.09}$ for HD~6680 and an orbital period of 466.63 $^{+1.47}_{-1.28}$ days, a semi-amplitude value of 353.99$^{+13.63}_{-23.04}$ m s$^{-1}$, and eccentricity of 0.36 $^{+0.06}_{-0.11}$  for HD~112300.
Thus, we obtain a minimum companion mass of 28.26$^{+2.06}_{-2.18}$ $M_{J}$ and a semi-major axis of 2.03$^{+0.01}_{-0.01}$ AU for HD~6680 as well as a minimum companion mass of 15.83$^{+2.34}_{-2.74}$ $M_{J}$ and semi-major axis of 1.33 $^{+0.08}_{-0.11}$ AU for HD~112300.
The radius of the M giant HD~6860 is 86 $R_{\odot}$, which makes it, at the time of this study, the largest star with a sub-stellar companion.

After removing the RV signal attributed to the companions, the remaining RV variations are 123.7~m~s$^{-1}$ for HD~6860 and 93.6~m~s$^{-1}$ for HD~112300. The variations are larger than those expected for the stellar activity, given the errors of the measurements.
Additional periodic variations can be considered and duly weighed because the values are significantly larger than the RV precision
for a RV standard star $\tau$ Ceti (7.5~m~s$^{-1}$) and are larger than the typical internal error of an individual RV accuracy rate of $\sim$ 12~m~s$^{-1}$.   \citet{2005A&A...437..743H} demonstrated that a high RMS of residuals ($\sim$~51~m~s$^{-1}$) can be found in the K1 giant HD~13189, showing significant short-term RV variability on the time scale of days that is most likely due to stellar oscillations.
\citet{2013A&A...549A...2L} showed that the RMS values of the RV residuals of the M2 III giant HD~220074 have a median value of 57~m s~$^{-1}$. Recently, \citet{2017ApJ...844...36L} found that the RMS values of the RV residuals of the M1 III giant HD~52030 and M0.5 III HD~208742 have  medians of 82~m s~$^{-1}$ and 55~m s~$^{-1}$, respectively.
Such behavior is typical for K and M giant stars, whether they are periodic or not, tending to increase toward later spectral types.

It is very meaningful and rare in itself that sub-stellar companions were discovered through RV observations around M (super) giants with a greatly expanded and active stellar atmosphere. In particular, the fact that sub-stellar companions survived RGB and AGB upheaval events presents another viable research topic related to the problem of planet survival.
\citet{2023Natur.618..917H} recently carried out follow-up observations and analysis on red giant 8 UMi, that had been an exoplanet system  in the RGB stage \citep{2015A&A...584A..79L}. They show that 8 UMi b have survived the RGB phase  and such case is expected to be common.  The system shows that a core-helium burning red giant can accommodate close planets and provides evidence for the role of non-standard stellar evolution in the extended survival of late-stage exoplanet systems. 
To understand such late evolutionary stage of exoplanet systems like those in this work will certainly become one of the main topics in future research on exoplanets.

\begin{acknowledgements}
BCL acknowledges partial support by the KASI (Korea Astronomy and Space Science Institute) grant
2023-1-832-03 and acknowledge support by the National Research Foundation of Korea(NRF) grant funded by the Korea government(MSIT) (No.2021R1A2C1009501).
MGP was supported by the Basic Science Research Program through the National Research Foundation of Korea (NRF) funded by the Ministry of Education (2019R1I1A3A02062242) and KASI under the R\&D program supervised by the Ministry of Science, ICT and Future Planning.
HYC was supported by a National Research Foundation of Korea Grant funded by the Korean government (NRF-2018R1D1A3B070421880) and Basic Science Research Program through the NRF of Korea funded by the Ministry of Science, ICT and Future Planning (No. 2018R1A6A1A06024970).
BL and JRK  acknowledge support by the NRF grant funded by MSIT (Grant No. 2022R1C1C2004102)
 This research made use of the SIMBAD database, operated at the CDS, Strasbourg, France.
\end{acknowledgements}


\begin{appendix} 
\section{RV measurements}
In this appendix we present all observational data collected with
the BOES. We list the observation dates Julian date (JD), the radial velocities (RV), and uncertainly ($\pm \sigma$), respective.

\begin{table*}
\renewcommand{\thetable}{\arabic{table}}
\centering
\caption{RV measurements for HD 6680 from December 2005 to December 2021 using the BOES.} \label{tab:rv1}
\begin{tabular}{ccccccccc}
\hline
\hline
JD & RV  & $\pm \sigma$ & JD & RV  & $\pm \sigma$  & JD & RV  & $\pm \sigma$\\
$-$2,450,000 &{m\,s$^{-1}$}& {m\,s$^{-1}$}& $-$2,450,000 & {m\,s$^{-1}$} & {m\,s$^{-1}$} & $-$2,450,000 & {m\,s$^{-1}$} & {m\,s$^{-1}$}\\
\hline

3730.088502 &    -156.4   &     6.7 &   4393.170102 &    -250.1   &     6.2  &   6204.127930 &    -135.8   &     4.4  \\
3752.937657 &      45.3   &     6.3 &   4469.999819 &     115.1   &     5.5  &   6618.996453 &     399.3   &     6.6  \\
3757.939853 &     107.4   &     5.5 &   4536.926076 &     485.3   &     9.1  &   6618.997645 &     395.1   &     5.8  \\
3757.945362 &     104.2   &     5.3 &   4719.096662 &    -171.5   &     5.8  &   7424.910021 &    -117.2   &    25.4  \\
3761.934804 &      95.9   &     4.9 &   4719.101408 &    -170.4   &     6.9  &   7424.912208 &    -124.6   &     9.4  \\
3761.946446 &      99.9   &     4.6 &   4752.054746 &     -85.5   &     6.8  &   7757.039726 &     -16.1   &     5.8  \\
3981.156599 &     431.3   &     7.7 &   4752.058334 &     -91.3   &     6.9  &   7757.041022 &     -18.6   &     6.5  \\
3981.160720 &     434.9   &     5.8 &   4756.146417 &     128.0   &     7.5  &   7812.953663 &     263.2   &     6.9  \\
4007.071779 &     356.9   &     7.2 &   4825.061892 &    -271.1   &     5.3  &   7812.955254 &     268.7   &     6.7  \\
4007.077231 &     357.3   &     4.6 &   4849.019378 &    -191.1   &     6.0  &   8092.018429 &      -6.6   &     7.6  \\
4027.176493 &     191.5   &     5.5 &   4995.289415 &    -336.9   &     6.1  &   9572.086456 &    -318.4   &     7.4  \\
4122.946679 &     116.5   &     6.8 &   5170.962422 &     -72.3   &     5.8  &   9577.044732 &    -311.1   &    10.9  \\
4165.925882 &    -355.4   &     6.7 &   5456.014304 &      92.5   &     5.5  &   9577.044732 &    -300.8   &     9.9  \\
4166.914685 &    -344.1   &     5.4 &   5554.033385 &     -22.4   &     6.2  &   9577.044732 &    -306.7   &    10.5  \\
4382.273629 &    -180.0   &     9.6 &   6204.123926 &    -134.9   &     4.7  &               &             &          \\

\hline
\end{tabular}
\end{table*}

\begin{table*}
\renewcommand{\thetable}{\arabic{table}}
\centering
\caption{RV measurements for HD 18884 from October 2006 to November 2020 using the BOES.} \label{tab:rv2}
\begin{tabular}{ccccccccc}
\hline
\hline
JD & RV  & $\pm \sigma$ & JD & RV  & $\pm \sigma$  & JD & RV  & $\pm \sigma$\\
$-$2,450,000 &{m\,s$^{-1}$}& {m\,s$^{-1}$}& $-$2,450,000 & {m\,s$^{-1}$} & {m\,s$^{-1}$} & $-$2,450,000 & {m\,s$^{-1}$} & {m\,s$^{-1}$}\\
\hline
4008.327874 &    -416.1  &      8.9  &  6241.019934  &    360.0   &     8.2  &    7090.924128  &     98.6   &    12.9   \\
4027.187785 &    -342.2  &      9.9  &  6241.022851  &    346.8   &    10.6  &    7090.926593  &     87.5   &    11.2   \\
4122.953808 &    -357.8  &      9.0  &  6250.086207  &    -61.8   &     9.5  &    7298.144581  &    151.8   &    11.3   \\
4165.932315 &    -254.8  &      9.1  &  6250.087931  &    -79.5   &     8.3  &    7327.191041  &    216.1   &    10.4   \\
4166.921063 &    -308.4  &      8.3  &  6250.089760  &    -92.3   &     8.8  &    7327.192407  &    218.0   &    10.5   \\
4382.278706 &     398.9  &      9.7  &  6250.090963  &    -86.9   &     8.6  &    7327.193934  &    217.4   &    10.4   \\
4470.006168 &     167.6  &      9.0  &  6250.092167  &    -93.2   &     8.4  &    7412.978762  &     69.5   &     9.8   \\
4472.043204 &     311.0  &     10.4  &  6260.123972  &     97.0   &     9.0  &    7412.979954  &     72.0   &    10.7   \\
4535.935158 &       5.9  &      9.2  &  6260.125963  &     91.7   &     9.6  &    7424.019660  &    244.2   &     9.7   \\
4739.206735 &    -194.3  &      8.8  &  6288.049216  &    -34.3   &     8.8  &    7424.021743  &    240.1   &     9.8   \\
4752.184935 &     -43.3  &      9.6  &  6288.050593  &    -35.6   &     8.1  &    7705.176836  &   -325.2   &    10.9   \\
4756.127346 &    -163.4  &     17.0  &  6619.001056  &   -142.7   &    10.4  &    7757.045042  &   -111.0   &     9.8   \\
4756.136131 &    -154.0  &     10.5  &  6619.002942  &   -139.2   &    11.2  &    7812.958900  &   -176.3   &    10.2   \\
4825.068323 &    -137.5  &     10.0  &  6922.326324  &     62.7   &    11.9  &    8015.256720  &    232.1   &    10.4   \\
4849.025488 &    -300.6  &      9.4  &  6964.113858  &    232.8   &    14.0  &    8092.024752  &   -107.4   &     9.9   \\
5171.134698 &      19.7  &      9.6  &  6964.118036  &    235.4   &    10.7  &    8147.989780  &     58.6   &     9.6   \\
5248.974417 &     229.3  &      9.9  &  6964.123453  &    233.0   &    11.6  &    8386.162507  &   -429.2   &    11.5   \\
5456.328873 &     190.5  &      9.8  &  6964.128789  &    243.6   &    12.8  &    8416.041949  &   -575.7   &    25.2   \\
5554.037869 &      37.9  &      9.2  &  6972.147580  &     -5.4   &    10.6  &    8422.154571  &   -614.2   &    11.7   \\
5581.068317 &    -296.5  &      9.7  &  6972.149570  &     -6.2   &    11.0  &    8451.221566  &   -467.9   &    12.3   \\
6173.275641 &     265.3  &      9.8  &  6972.151434  &      6.2   &     9.9  &    8472.932541  &   -258.8   &    10.7   \\
6204.243142 &     293.8  &      9.9  &  7024.879488  &    -23.0   &     9.1  &    9161.160688  &     39.0   &    11.1   \\
6209.105254 &     295.6  &     10.1  &  7024.881375  &    -16.0   &     8.7  &    9161.163246  &     28.8   &    11.4   \\
6241.014471 &     338.5  &     10.5  &  7024.883736  &    -27.7   &     9.2  &                 &            &           \\
6241.017342 &     345.5  &     10.5  &  7090.922392  &     95.3   &    11.7  &                 &            &           \\

\hline
\end{tabular}
\end{table*}

\begin{table*}
\renewcommand{\thetable}{\arabic{table}}
\centering
\caption{RV measurements for HD 39801 from January 2008 to April 2021 using the BOES.} \label{tab:rv3}
\begin{tabular}{ccccccccc}
\hline
\hline
JD & RV  & $\pm \sigma$ & JD & RV  & $\pm \sigma$  & JD & RV  & $\pm \sigma$\\
$-$2,450,000 &{km\,s$^{-1}$}& {m\,s$^{-1}$}& $-$2,450,000 & {km\,s$^{-1}$} & {m\,s$^{-1}$} & $-$2,450,000 & {km\,s$^{-1}$} & {m\,s$^{-1}$}\\
\hline

4470.013168 &   -0.5649 &    0.0448 &   6377.001369 &   -5.3753  &   0.0430 &   7327.194710 &    1.6930  &   0.0548   \\
4470.014371 &   -0.6077 &    0.0437 &   6579.266762 &   -1.4754  &   0.0318 &   7327.195404 &    1.6694  &   0.0535   \\
4472.052617 &   -0.7368 &    0.0464 &   6579.267526 &   -1.4843  &   0.0295 &   7327.196099 &    1.6765  &   0.0561   \\
4472.150716 &   -0.7316 &    0.0456 &   6579.268255 &   -1.4498  &   0.0309 &   7412.986609 &    1.7131  &   0.0394   \\
4505.106474 &   -0.9494 &    0.0266 &   6620.075080 &   -1.5220  &   0.0338 &   7412.987535 &    1.7129  &   0.0467   \\
4507.013924 &   -0.9456 &    0.0276 &   6620.075983 &   -1.5171  &   0.0353 &   7424.029588 &    1.8087  &   0.0428   \\
4502.347253 &    0.2657 &    0.0272 &   6679.109560 &   -2.7872  &   0.0344 &   7424.030965 &    1.7969  &   0.0394   \\
4507.019155 &   -0.9959 &    0.0259 &   6679.110300 &   -2.7478  &   0.0316 &   7429.141308 &    1.9807  &   0.0413   \\
4507.021331 &   -0.9795 &    0.0253 &   6679.111030 &   -2.7768  &   0.0330 &   7429.142442 &    1.9896  &   0.0432   \\
4535.946989 &   -1.5355 &    0.0336 &   6739.926676 &   -4.1461  &   0.0283 &   7475.012750 &    1.7012  &   0.0597   \\
4537.017802 &   -1.5644 &    0.0353 &   6739.927469 &   -4.1588  &   0.0261 &   7475.014086 &    1.7590  &   0.0359   \\
4824.169284 &    3.1342 &    0.0721 &   6922.326715 &   -0.4178  &   0.0284 &   7705.178234 &   -0.0925  &   0.0409   \\
4824.170094 &    3.1987 &    0.0739 &   6960.309345 &    0.6122  &   0.0769 &   7705.179299 &   -0.0477  &   0.0400   \\
4849.032189 &    3.7155 &    0.0621 &   6960.310202 &    0.6189  &   0.0619 &   7757.067799 &    0.6623  &   0.0399   \\
4849.033289 &    3.7346 &    0.0657 &   6964.148528 &    0.6304  &   0.0379 &   7812.967220 &   -0.6463  &   0.0352   \\
4881.053413 &    4.4718 &    0.0512 &   6964.149951 &    0.5745  &   0.0439 &   8015.256744 &   -2.6382  &   0.0652   \\
4930.982267 &    2.9147 &    0.0350 &   6964.151109 &    0.5896  &   0.0417 &   8092.184677 &   -0.4550  &   0.0592   \\
5171.130641 &    1.5919 &    0.0429 &   6964.152266 &    0.6269  &   0.0433 &   8092.185702 &   -0.4148  &   0.0508   \\
5248.982285 &    2.7903 &    0.0942 &   6972.153102 &    0.7037  &   0.0387 &   8109.242300 &   -0.6397  &   0.0602   \\
5248.983072 &    2.8145 &    0.0845 &   6972.154144 &    0.7026  &   0.0390 &   8109.251721 &   -0.5684  &   0.0402   \\
5581.119825 &   -0.5549 &    0.0756 &   6972.154972 &    0.6867  &   0.0366 &   8109.252277 &   -0.6093  &   0.0450   \\
6023.996060 &   -3.5320 &    0.0810 &   6972.155846 &    0.6945  &   0.0389 &   8109.252844 &   -0.5899  &   0.0437   \\
6023.995064 &   -3.5004 &    0.0877 &   6972.156436 &    0.7029  &   0.0368 &   8422.156652 &   -3.8636  &   0.0686   \\
6173.286187 &   -1.7249 &    0.0139 &   6972.157026 &    0.6769  &   0.0370 &   9161.277398 &    2.3194  &   0.0771   \\
6210.216620 &   -1.6074 &    0.0230 &   7025.008976 &    1.5985  &   0.0366 &   9161.278510 &    2.2810  &   0.0527   \\
6210.217418 &   -1.6201 &    0.0242 &   7025.009833 &    1.6076  &   0.0350 &   9161.280292 &    2.3451  &   0.0661   \\
6210.218182 &   -1.5968 &    0.0244 &   7025.010677 &    1.6268  &   0.0326 &   9296.025984 &    1.8662  &   0.0560   \\
6210.218911 &   -1.5898 &    0.0225 &   7090.933902 &    0.7222  &   0.0363 &   9310.916041 &    2.2927  &   0.0942   \\
6348.116361 &   -4.6275 &    0.0619 &   7090.935106 &    0.7298  &   0.0354 &   9310.915347 &    2.1380  &   0.1242   \\
6348.117947 &   -4.6233 &    0.0580 &   7090.936298 &    0.7500  &   0.0355 &   9311.956918 &    2.3534  &   0.1173   \\
6377.001369 &   -5.3753 &    0.0430 &   7301.272827 &    1.1379  &   0.0749 &               &            &            \\

\hline
\end{tabular}
\end{table*}

\begin{table*}
\renewcommand{\thetable}{\arabic{table}}
\centering
\caption{RV measurements for HD 42995 from October 2006 to April 2021 using the BOES.} \label{tab:rv4}
\begin{tabular}{ccccccccc}
\hline
\hline
JD & RV  & $\pm \sigma$ & JD & RV  & $\pm \sigma$  & JD & RV  & $\pm \sigma$\\
$-$2,450,000 &{km\,s$^{-1}$}& {m\,s$^{-1}$}& $-$2,450,000 & {km\,s$^{-1}$} & {m\,s$^{-1}$} & $-$2,450,000 & {km\,s$^{-1}$} & {m\,s$^{-1}$}\\
\hline

4018.306232 &   -2.0091  &   0.1104  &   5555.285963 &    1.8123 &    0.0453  &   7812.970371 &    3.0843 &    0.0247   \\
4118.340291 &    1.4982  &   0.0164  &   5581.122792 &    1.8923 &    0.0322  &   8015.258807 &    2.3274 &    0.0442   \\
4126.132301 &   -0.5318  &   0.0101  &   6024.000160 &   -2.4481 &    0.0263  &   8092.191228 &    2.8088 &    0.0246   \\
4147.087800 &    0.4856  &   0.0213  &   6210.314520 &   -9.9701 &    0.0241  &   8109.245415 &    3.2853 &    0.0272   \\
4389.362718 &    1.8351  &   0.0367  &   6620.078504 &  -10.3738 &    0.0284  &   8109.247498 &    3.3102 &    0.0284   \\
4470.018469 &    1.9663  &   0.0809  &   6620.080946 &  -10.3429 &    0.0293  &   8109.248633 &    3.3662 &    0.0312   \\
4472.157305 &    1.4009  &   0.0729  &   7424.035619 &    2.3596 &    0.0271  &   8109.249778 &    3.3598 &    0.0320   \\
4505.110781 &    0.9326  &   0.0464  &   7424.038918 &    2.4064 &    0.0246  &   8233.976097 &    1.7576 &    0.0484   \\
4537.024184 &    1.4670  &   0.0289  &   7424.042899 &    2.4238 &    0.0271  &   8233.978840 &    1.6883 &    0.0849   \\
4756.238517 &    2.0810  &   0.0389  &   7429.147157 &    2.0440 &    0.0254  &   8473.001639 &    1.6824 &    0.0356   \\
4824.173877 &    2.2273  &   0.0169  &   7429.150617 &    2.0615 &    0.0251  &   8516.903807 &    2.2192 &    0.0245   \\
4849.037301 &    2.6326  &   0.0211  &   7475.018895 &    2.7767 &    0.0713  &   8577.000884 &    1.9320 &    0.0301   \\
4881.059476 &    2.8058  &   0.0251  &   7497.953461 &    2.5630 &    0.0955  &   9161.287625 &   -8.2797 &    0.0659   \\
5171.141965 &    2.2005  &   0.1054  &   7500.946357 &    2.2119 &    0.0953  &   9296.033251 &  -13.9175 &    0.0261   \\
5248.986186 &    2.7100  &   0.0204  &   7500.949840 &    2.2121 &    0.0930  &   9310.916432 &  -13.6461 &    0.0365   \\
5456.337385 &    2.1245  &   0.0372  &   7705.183231 &    3.2439 &    0.0862  &   9310.917127 &  -13.6773 &    0.0357   \\

\hline
\end{tabular}
\end{table*}

\begin{table*}
\renewcommand{\thetable}{\arabic{table}}
\centering
\caption{RV measurements for HD 44478 from October 2006 to April 2021 using the BOES.} \label{tab:rv5}
\begin{tabular}{ccccccccc}
\hline
\hline
JD & RV  & $\pm \sigma$ & JD & RV  & $\pm \sigma$  & JD & RV  & $\pm \sigma$\\
$-$2,450,000 &{km\,s$^{-1}$}& {m\,s$^{-1}$}& $-$2,450,000 & {km\,s$^{-1}$} & {m\,s$^{-1}$} & $-$2,450,000 & {km\,s$^{-1}$} & {m\,s$^{-1}$}\\
\hline
3729.266531 &   -0.4567  &   0.0207  &  4125.107539  &  -0.4979 &    0.0061 &    5249.150417 &    0.1572  &   0.0089  \\
3730.114809 &   -0.4126  &   0.0227  &  4147.092587  &   0.2557 &    0.0110 &    5456.339492 &   -0.1766  &   0.0096  \\
3730.129091 &   -0.4298  &   0.0232  &  4147.096047  &   0.2536 &    0.0097 &    5555.288285 &    0.3259  &   0.0513  \\
3751.195164 &   -0.1471  &   0.0117  &  4165.959846  &  -0.6806 &    0.0138 &    5581.125497 &    0.0333  &   0.0077  \\
3756.196836 &   -0.2940  &   0.0079  &  4204.956383  &  -0.2667 &    0.0089 &    6024.004140 &   -0.4092  &   0.0082  \\
3758.081788 &   -0.2780  &   0.0077  &  4213.989839  &  -0.7607 &    0.0151 &    6210.317068 &   -0.0737  &   0.0250  \\
3763.050676 &   -0.3207  &   0.0094  &  4389.367366  &   1.0024 &    0.0208 &    6620.083125 &    0.6108  &   0.0134  \\
3808.988104 &   -0.4554  &   0.0143  &  4458.316078  &   0.2718 &    0.0174 &    6620.085081 &    0.6098  &   0.0160  \\
3823.975127 &    0.2742  &   0.0096  &  4470.022988  &   1.6165 &    0.0111 &    7424.048600 &   -0.0054  &   0.0121  \\
3833.968669 &    0.1674  &   0.0092  &  4472.165026  &   1.6783 &    0.0118 &    7424.051875 &   -0.0111  &   0.0129  \\
4008.340565 &    0.0409  &   0.0130  &  4483.154563  &  -0.0315 &    0.0183 &    7429.154848 &    0.2138  &   0.0127  \\
4008.352938 &    0.0407  &   0.0115  &  4505.115691  &   0.7210 &    0.0089 &    7429.157684 &    0.2062  &   0.0138  \\
4018.310933 &   -0.0442  &   0.0091  &  4537.030315  &  -0.0949 &    0.0230 &    7475.023306 &   -0.3539  &   0.0093  \\
4027.274381 &   -0.8644  &   0.0143  &  4756.245320  &  -1.0386 &    0.0244 &    7705.187279 &    0.1471  &   0.0086  \\
4038.392748 &   -0.0701  &   0.0107  &  4824.177863  &  -0.0259 &    0.0082 &    7757.073413 &    0.3996  &   0.0120  \\
4051.385285 &   -0.4293  &   0.0104  &  4849.042211  &   0.1732 &    0.0081 &    7812.975540 &    1.0795  &   0.0133  \\
4110.974296 &   -0.6975  &   0.0168  &  4881.067014  &   0.3508 &    0.0076 &    8092.197386 &   -0.5186  &   0.0107  \\
4122.993497 &   -0.2784  &   0.0063  &  4930.046764  &   0.1560 &    0.0107 &    9299.001825 &   -0.4132  &   0.0153  \\
4123.000264 &   -0.2757  &   0.0052  &  5171.146912  &   0.0258 &    0.0125 &                &            &           \\
\hline
\end{tabular}
\end{table*}

\begin{table*}
\renewcommand{\thetable}{\arabic{table}}
\centering
\caption{RV measurements for HD 112300 from June 2006 to December 2021 using the BOES.} \label{tab:rv6}
\begin{tabular}{ccccccccc}
\hline
\hline
JD & RV  & $\pm \sigma$ & JD & RV  & $\pm \sigma$  & JD & RV  & $\pm \sigma$\\
$-$2,450,000 &{m\,s$^{-1}$}& {m\,s$^{-1}$}& $-$2,450,000 & {m\,s$^{-1}$} & {m\,s$^{-1}$} & $-$2,450,000 & {m\,s$^{-1}$} & {m\,s$^{-1}$}\\
\hline

3899.050803  &   -378.0   &    13.6 &   5171.382920  &   -284.0   &    17.0 &   7475.036828  &    -58.5 &      16.1  \\
3899.056115  &   -377.4   &    13.2 &   5248.204644  &   -244.5   &    15.3 &   7525.054240  &   -239.1 &      15.7  \\
4147.311993  &     28.3   &    15.8 &   5356.997152  &   -352.5   &    14.1 &   8092.335474  &   -160.6 &      16.3  \\
4458.417618  &   -225.9   &    15.6 &   5554.376531  &    230.7   &    20.2 &   8234.081431  &    218.5 &      15.3  \\
4470.414622  &   -268.3   &    16.2 &   5672.153955  &      5.2   &    17.5 &   9330.050431  &    241.2 &      21.4  \\
4505.242519  &   -230.4   &    16.4 &   6024.231099  &     77.0   &    16.7 &   9330.967062  &    350.6 &      15.5  \\
4516.268428  &    184.5   &    16.0 &   7429.173250  &    183.4   &    19.7 &   9330.048695  &    288.8 &     104.6  \\
4537.175405  &    261.5   &    15.5 &   7429.177313  &    158.5   &    18.6 &   9552.332309  &      6.7 &      15.0  \\
4619.020239  &    280.3   &    15.4 &   7429.181283  &    183.6   &    20.5 &   9552.332309  &      9.2 &      14.1  \\
4847.385270  &   -227.5   &    14.8 &   7469.122485  &    147.4   &    16.6 &   9552.332309  &     -4.5 &      13.4  \\
4930.153182  &   -131.5   &    15.4 &   7469.125471  &    145.5   &    15.7 &   9561.378265  &    -32.8 &      16.5  \\
4971.051757  &     75.2   &    16.6 &   7469.129730  &    139.4   &    14.5 &                &          &            \\

\hline
\end{tabular}
\end{table*}

\begin{table*}
\renewcommand{\thetable}{\arabic{table}}
\centering
\caption{RV measurements for HD 146501 from June 2006 to March 2021 using the BOES.} \label{tab:rv7}
\begin{tabular}{ccccccccc}
\hline
\hline
JD & RV  & $\pm \sigma$ & JD & RV  & $\pm \sigma$  & JD & RV  & $\pm \sigma$\\
$-$2,450,000 &{m\,s$^{-1}$}& {m\,s$^{-1}$}& $-$2,450,000 & {m\,s$^{-1}$} & {m\,s$^{-1}$} & $-$2,450,000 & {m\,s$^{-1}$} & {m\,s$^{-1}$}\\
\hline

3899.086682  &     -6.2  &      8.8 &   5672.158204 &     -26.4  &     10.7  &   6801.037418 &     -68.6 &      13.1  \\
4151.372718  &    139.1  &     10.0 &   6024.234129 &     125.4  &     10.1  &   6823.001466 &    -184.4 &      10.2  \\
4263.101716  &    136.6  &      9.8 &   6711.388187 &    -118.5  &      9.9  &   6823.003711 &    -185.7 &      11.0  \\
4506.386631  &     87.3  &      9.4 &   6711.390641 &    -114.3  &     10.1  &   6823.005945 &    -191.0 &      10.0  \\
4931.185126  &    142.3  &     10.8 &   6711.392088 &    -113.6  &     10.8  &   6827.147000 &     125.3 &      10.3  \\
4970.161001  &     33.3  &     10.3 &   6711.392991 &    -109.4  &     11.0  &   6827.149396 &     123.6 &      10.3  \\
4970.179809  &     31.4  &     11.7 &   6711.393836 &    -112.4  &     11.4  &   6827.151433 &     126.1 &      10.4  \\
4970.181696  &     37.3  &     10.1 &   6711.394681 &    -113.8  &     10.8  &   6827.152995 &     129.5 &      10.3  \\
4971.239922  &     64.3  &     11.3 &   6711.395526 &    -110.5  &     11.8  &   6920.933931 &     119.9 &       9.9  \\
4995.051144  &    -46.2  &     12.7 &   6740.249523 &     -41.6  &     10.1  &   7169.223837 &    -101.3 &      10.1  \\
5248.373658  &   -142.0  &     10.0 &   6790.159034 &      62.4  &     11.6  &   7475.231317 &     101.3 &       9.8  \\
5357.004119  &    -21.6  &      9.2 &   6790.161187 &      67.9  &     10.7  &   7525.132860 &     124.2 &      10.1  \\
5455.953732  &     99.2  &     10.6 &   6790.163074 &      60.2  &     12.0  &   9299.214768 &    -129.1 &      10.5  \\

\hline
\end{tabular}
\end{table*}

\begin{table*}
\renewcommand{\thetable}{\arabic{table}}
\centering
\caption{RV measurements for HD 156014 from June 2006 to November 2016 using the BOES.} \label{tab:rv8}
\begin{tabular}{ccccccccc}
\hline
\hline
JD & RV  & $\pm \sigma$ & JD & RV  & $\pm \sigma$  & JD & RV  & $\pm \sigma$\\
$-$2,450,000 &{km\,s$^{-1}$}& {m\,s$^{-1}$}& $-$2,450,000 & {km\,s$^{-1}$} & {m\,s$^{-1}$} & $-$2,450,000 & {km\,s$^{-1}$} & {m\,s$^{-1}$}\\
\hline

3899.081654 &   -0.2412  &   0.0133 &    4971.242883 &   -1.0731  &   0.0445 &    7475.238397 &   -1.5142 &    0.1456  \\
4017.945815 &    0.0795  &   0.0214 &    4995.040650 &   -0.2860  &   0.0326 &    7525.136765 &    0.6087 &    0.0767  \\
4262.222066 &   -0.1772  &   0.0299 &    5248.374783 &   -0.2800  &   0.0507 &    7525.139346 &    0.4693 &    0.0702  \\
4263.111021 &   -0.1183  &   0.0298 &    5357.006520 &   -1.8538  &   0.0532 &    7525.242555 &    0.5366 &    0.0707  \\
4506.389744 &   -1.0864  &   0.0487 &    5455.995217 &   -0.1344  &   0.0661 &    7525.244129 &    0.3477 &    0.0685  \\
4619.240943 &    0.5887  &   0.0513 &    5672.161343 &    1.0211  &   0.0617 &    7704.882731 &    0.6466 &    0.0885  \\
4750.944989 &    0.7558  &   0.0267 &    6024.257566 &    2.7067  &   0.0722 &    7704.885995 &    0.5886 &    0.0293  \\
4970.168132 &   -1.0746  &   0.0653 &    6619.871127 &    1.3102  &   0.1279 &    7704.885995 &    0.5886 &    0.0293  \\
4970.172173 &   -1.0857  &   0.0683 &    7475.234728 &   -1.3237  &   0.1543 &                &           &            \\

\hline
\end{tabular}
\end{table*}

\begin{table*}
\renewcommand{\thetable}{\arabic{table}}
\centering
\caption{RV measurements for HD 183030 from June 2010 to January 2022 using the BOES.} \label{tab:rv9}
\begin{tabular}{ccccccccc}
\hline
\hline
JD & RV  & $\pm \sigma$ & JD & RV  & $\pm \sigma$  & JD & RV  & $\pm \sigma$\\
$-$2,450,000 &{m\,s$^{-1}$}& {m\,s$^{-1}$}& $-$2,450,000 & {m\,s$^{-1}$} & {m\,s$^{-1}$} & $-$2,450,000 & {m\,s$^{-1}$} & {m\,s$^{-1}$}\\
\hline

5357.175344 &    -339.1  &     12.3 &   7066.254678 &     191.1  &     20.4 &   7895.099207 &     386.2  &     12.7  \\
5455.227744 &    -112.1  &     10.4 &   7068.219814 &     204.8  &     13.8 &   8011.015087 &     -15.3  &     14.3  \\
5664.138335 &      84.7  &     11.7 &   7148.103251 &    -209.5  &     12.4 &   8091.960588 &    -449.8  &     11.9  \\
5843.018018 &      -2.5  &     10.8 &   7169.282932 &    -531.2  &     16.2 &   8109.934247 &    -707.6  &     11.8  \\
6409.108541 &      78.9  &      8.4 &   7300.969354 &     -39.2  &     14.6 &   8519.174594 &     -36.7  &     11.5  \\
6710.334177 &     106.3  &     10.2 &   7475.198784 &     185.3  &     10.9 &   8562.164467 &     281.6  &     12.1  \\
6739.178024 &    -328.2  &     11.7 &   7527.123920 &       2.4  &     16.2 &   8743.970114 &    -483.2  &     15.4  \\
6805.195572 &     -60.7  &     14.0 &   7704.958298 &    -119.4  &     11.4 &   8772.972615 &     -91.6  &     10.7  \\
6920.057371 &     411.5  &      9.5 &   7820.054643 &     218.3  &     14.1 &   9580.967424 &      46.6  &     63.9  \\
6970.891866 &     525.6  &     13.6 &   7856.262548 &     580.7  &     19.5 &   9580.967424 &     223.3  &     66.6  \\

\hline
\end{tabular}
\end{table*}


\section{CORNER PLOTS}
We present Posterior distributions of the parameters sampled from the model MCMC fit. 
\begin{figure*}
 \centering
   \includegraphics[width=18.5cm]{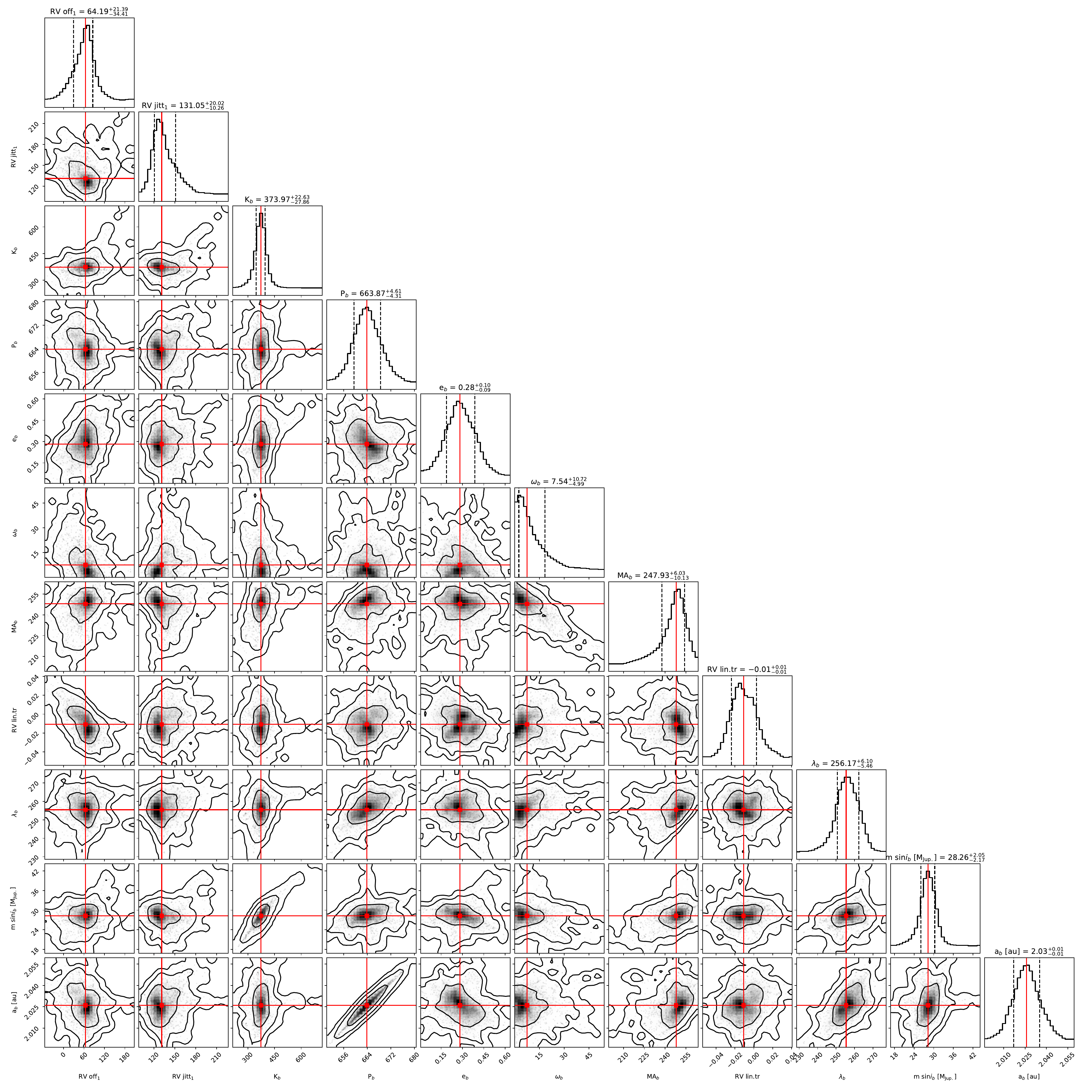}
      \caption{Corner plot showing the posterior distributions of the main internal structure parameters of HD 6860 b.  The titles of each column correspond to the median and 1 $\sigma$, which are also shown with the dashed lines. It displaying the results of MCMC analysis on the corrected RV time-series is presented. This analysis involves fitting a model that consists of an offset, an RV jitter, the semi-amplitude ($K_b$), the orbital period ($P_b$), Orbital eccentricity ($e_b$), Periastron angle ($\omega_b$), Mean anomaly (MA$_b$), Slope (RV lin.tr), Mean longitude $\lambda_b$, Minimum mass ($m$ sin $i$), and Semi-major axis ($a_b$), respectively. }
         \label{f14_hd6860_mcmc} \end{figure*}
%

%
   \begin{figure*}
   \centering
   \includegraphics[width=18.5cm]{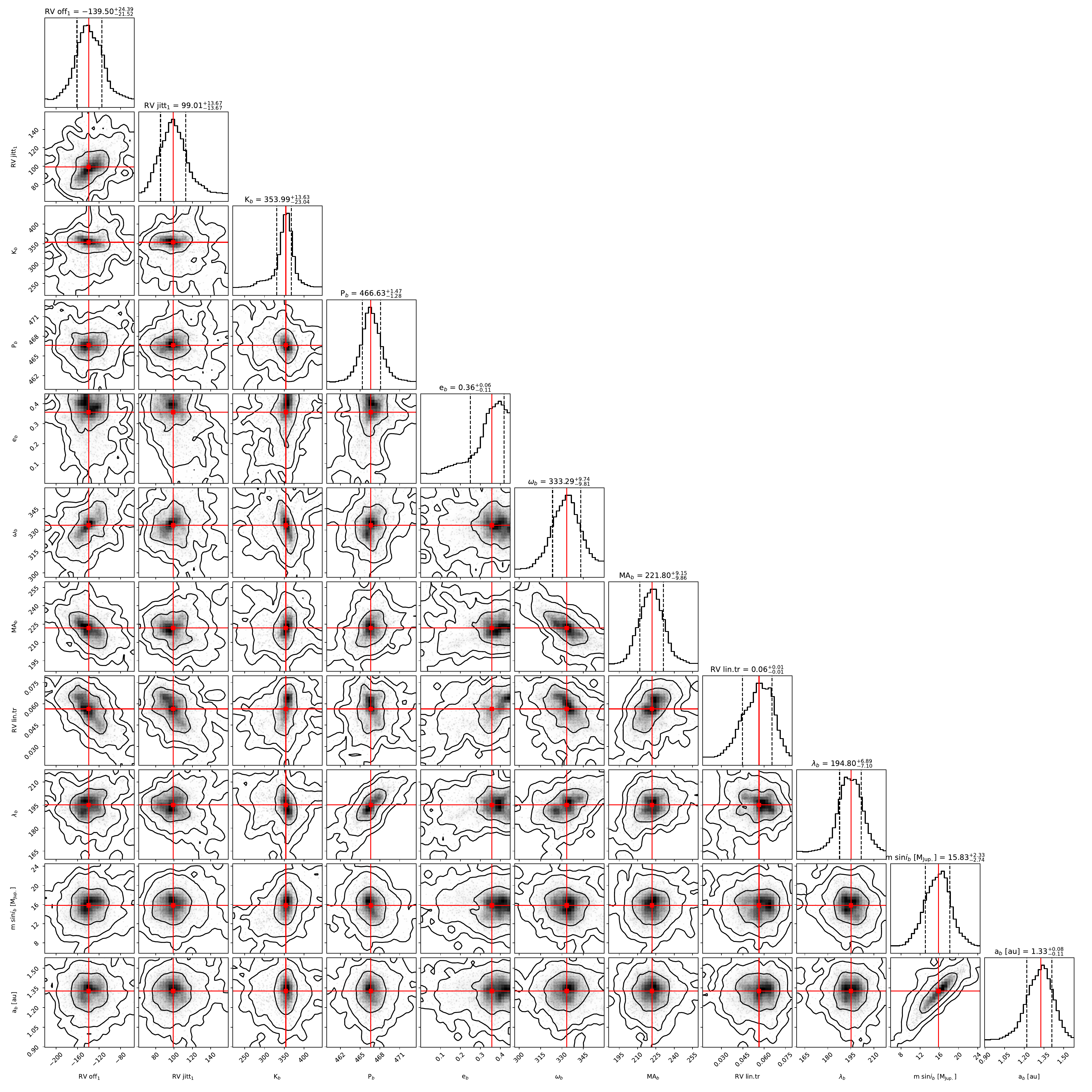}
      \caption{Corner plot showing the posterior distributions of the main internal structure parameters of HD 112300 b.  The titles of each column correspond to the median and 1 $\sigma$, which are also shown with the dashed lines. It displaying the results of MCMC analysis on the corrected RV time-series is presented. This analysis involves fitting a model that consists of an offset, an RV jitter, the semi-amplitude ($K_b$), the orbital period ($P_b$), Orbital eccentricity ($e_b$), Periastron angle ($\omega_b$), Mean anomaly (MA$_b$), Slope (RV lin.tr), Mean longitude $\lambda_b$, Minimum mass ($m$ sin $i$), and Semi-major axis ($a_b$), respectively.	
              }
         \label{f15_hd112300_mcmc}
   \end{figure*}

\end{appendix}

\end{document}